\documentclass[aps, prl, reprint, showpacs, twocolumn, 10pt, groupedaddress,noeprint]{revtex4-2}

\usepackage{graphicx, amsmath, dsfont, dcolumn, bm, times, xcolor, braket, amssymb,natbib}

\begin{document}

\title{Josephson quantum mechanics at odd parity}

\author{Manuel~Houzet, Julia~S.~Meyer}
\affiliation{Univ.~Grenoble Alpes, CEA, Grenoble INP, IRIG, {Pheliqs}, 38000 Grenoble, France}
\author{Yuli V. Nazarov}
\affiliation{Kavli Institute of Nanoscience, Delft University of Technology, 2628 CJ Delft, The Netherlands}

\date{\today}

\begin{abstract}
A Josephson junction may be in a stable odd parity state when a single quasiparticle is trapped in an Andreev bound state. Embedding such junction in an electromagnetic environment gives rise to a special quantum mechanics of superconducting phase that we investigate theoretically. Our analysis covers several representative cases, from the lifting of the supercurrent quench due to quasiparticle poisoning for a low ohmic impedance of the environment, to a Schmid transition in a current-biased junction that for odd parity occurs at four times bigger critical impedance. For intermediate impedances, the supercurrent in the odd state is higher than in the even one. 
\end{abstract}

\maketitle

The energy of a tunnel junction between two superconducting leads depends periodically on the difference of superconducting phases of the two, in short, on the phase. This is the celebrated Josephson effect~\cite{Josephson}: the phase dependence of this energy gives rise to a persistent superconducting current between the leads. Later, it has been understood that the phase becomes a quantum-fluctuating variable if a Josephson junction is embedded in an electromagnetic circuit~\cite{oldbook}.  Earlier studies concentrated on a dissipative electromagnetic environment and were essential for establishing the modern theory of dissipative quantum mechanics~\cite{LeggetReview,Weiss}. A highlight of this research was the prediction of the Schmid transition~\cite{Schmid}: the vanishing of the Josephson energy at a critical value of the circuit impedance $R$, $ 2 e^2 R/\pi \hbar \equiv \alpha =1$. While this prediction is theoretically indisputable, the controversy concerning its experimental verification~\cite{Murani,subero2022bolometric} 
may have been resolved recently~\cite{Kuzmin}. The further development of Josephson quantum mechanics evolved from dissipative circuits to dissipationless Coulomb islands. The resulting Josephson-based superconducting qubits \cite{schoen2001,blais2021circuit} are at the frontline of modern quantum technology applications.

 
There is something to add to this well-established field.
In fact, the Josephson energy is related to Andreev bound states (ABS) in the junction~\cite{BeenakkerABS}  and does depend on their occupation.  Of the two equal-weight superpositions with respect to the right/left leads in which a quasiparticle may be in, only one gives rise to a bound state. Owing to parity conservation in superconductors~\cite{AverinNazarov}, a state with a single quasiparticle trapped in the lowest ABS (the odd parity ground state) is stable despite having a bigger energy than the state without quasiparticles (the even parity state). Physically, the parity can only be relaxed if a stray quasiparticle from a lead comes to the junction and annihilates the trapped one. Since the concentration of the quasiparticles in the leads is  vanishingly small at low temperatures, the lifetime of the odd parity ground state is macroscopically long: lifetimes of several  minutes have been measured \cite{LongTime}. We note that a single quasiparticle trapped in a spin-degenerate ABS eventually quenches the contribution of this level to the Josephson energy: this is called the quasiparticle poisoning and has been observed in \cite{QuantronicsPoisoning}.
 When spin-degeneracy is lifted (in finite-length junctions with spin-orbit coupling), the stability of  these odd states provided the opportunity 
for a new kind of qubits: Andreev spin qubits, proposed in \cite{Chtchelkatchev2003,Padurariu2010} and realized in \cite{SpinQubitExp}.
 In recent years, there is an outburst of studies of ABS in superconducting nanostructures, including spectroscopically resolved ABS and odd parity ground states in a junction ~\cite{RamonReview}.

\begin{figure}
\includegraphics[width=\columnwidth]{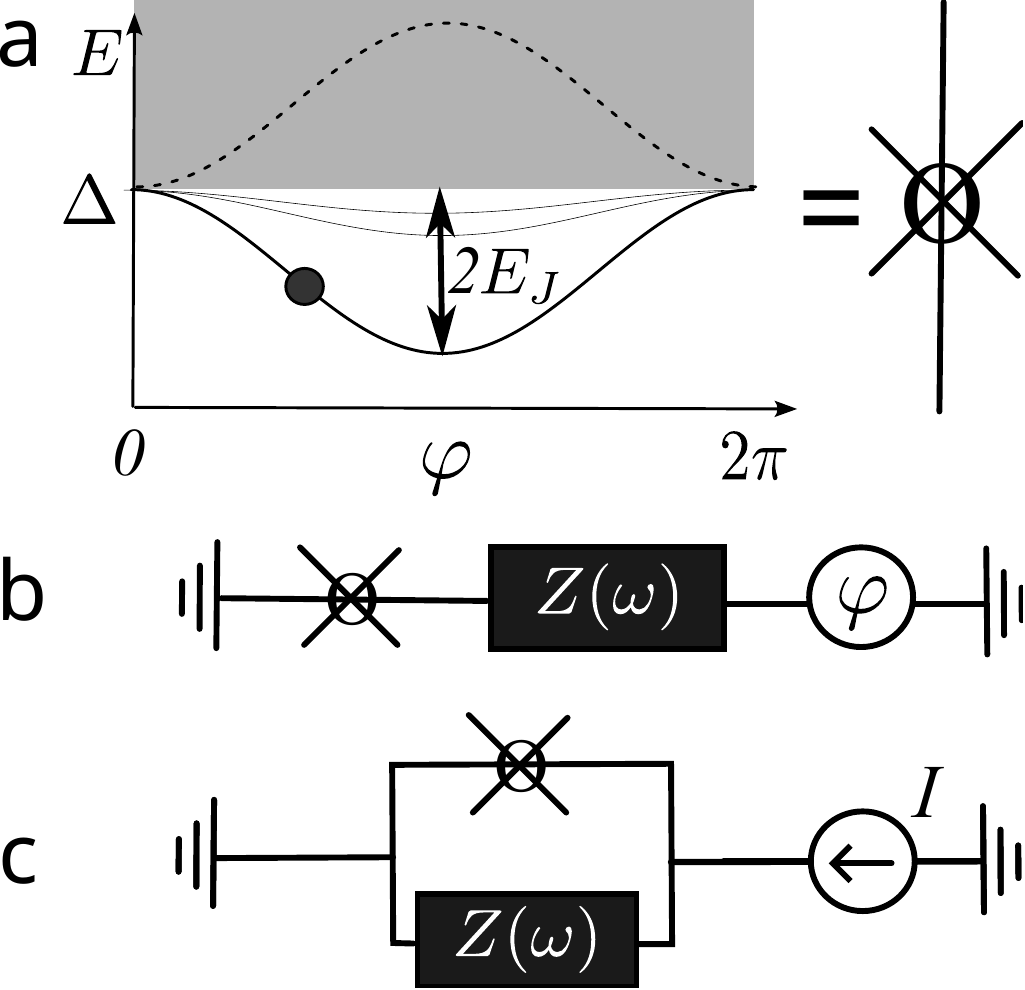}
\caption{\label{fig:FirstFigure}
a. The odd parity Josephson junction. A single quasiparticle is trapped in the lowest Andreev level separated by $2E_J \sin^2\frac{\varphi}{2} \ll \Delta$ from the edge of the continuous quasiparticle spectrum at the superconducting energy gap $\Delta$. In the bound state, the quasiparticle is in a certain superposition, $s=1$, the anti-bound state corresponding to $s=-1$ (dashed curve) belongs to the continuous spectrum.
b.-c. The Josephson quantum mechanics at odd parity: the odd parity Josephson junction is embedded in a linear electromagnetic environment with frequency-dependent impedance $Z(\omega)$ that causes quantum fluctuations of the phase.  b. and c. correspond to phase and current bias, respectively. }
\end{figure}

This makes it relevant to extend the Josephson quantum mechanics  
to the case of a circuit embedding a Josephson junction in the odd parity ground state. Such quantum mechanics at odd parity should be quite distinct from the conventional one. For instance, for a short single-channel junction, quasiparticle poisoning is expected to  completely quench the supercurrent~\cite{BeenakkerABS}. Thus a naive and, as we will see, wrong expectation is that the junction is not present in the circuit at all. In our pivotal study, we consider a tunnel junction where the ABS are close to the superconducting gap edge, disregard weak spin-orbit interaction, and mainly concentrate on  the instructive single-level, single-junction case, see Fig.~\ref{fig:FirstFigure}.

In this Letter, we provide a general description of Josephson quantum mechanics at odd parity revealing its intriguing mathematical structure. We also present the detailed analysis for three relevant cases. For low ohmic impedance, we demonstrate the  incompleteness of supercurrent quenching  and reveal a supercurrent jump at zero phase. For arbitrary ohmic impedance and {\em phase} bias, we establish a slower suppression of the Josephson energy in the odd state  than in the even one: the supercurrent in the odd state thus becomes {\it higher} than in the even one, both remaining finite at any $\alpha$ as already shown in the even case~\cite{Hekking1997}. At sufficiently large impedance, {\em both} right/left superpositions 
form a bound state. While their phase-dependence is suppressed upon increasing the impedance, their average binding energy tends to a constant. In addition to this, for arbitrary ohmic impedance and {\em current} bias, we encounter a Schmid transition at a higher value of the impedance than in the even state, namely, at $\alpha = 4$. The bound states persist for both superpositions and are {\it degenerate} for $\alpha >4$. These predictions can be tested in forthcoming experiments.

Let us sketch here the general derivation: all details are provided in~\footnote{\label{Supplemental} See Supplemental Material, which includes Refs.~\cite{LarkinDisorder,Feigelman}, at [URL will be inserted by publisher] for the details of the derivations.}. At even parity, the Hamiltonian describing a Josephson junction embedded in a general linear environment, see Fig.~\ref{fig:FirstFigure}b, reads \cite{CaldeiraLegget}
\begin{equation}
\label{Hamiltonian}
H_{{\rm e}} = H_{{\rm env}} - E^*_J \cos \hat \varphi,
\end{equation}
where $H_{{\rm env}}$ is a Hamiltonian of non-interacting bosons, the operator of the phase drop at the junction, $\hat \varphi$, consists  of  the phase bias $\varphi$ and a linear superposition of environmental bosons, and $E^*_J$ is the even-state Josephson energy. 
The coefficients in the superposition are chosen such as to reproduce the frequency-dependent impedance of the environment,  $Z(\omega)$.
An alternative description \cite{Schoen1990} employs a path integral over a variable $\varphi(\tau)$ defined in imaginary time. The action that defines the path weight reads
\begin{equation}
{\cal S} = \sum_\omega\frac{| \omega|}{8e^2 Z(i|\omega|)} |\varphi(\omega)|^2 - E^*_J \int d\tau \cos \varphi(\tau),
\end{equation}
$\varphi(\omega)$ being the Fourier transform of $\varphi(\tau)$.

To describe the odd parity situation, we first augment the Hilbert space with the states of a single quasiparticle to reduce it at a later stage of the derivation. Without the environment, this gives the binding energy $\Omega$, measured from the edge of the continuum,  in the following form:
\begin{equation}
\label{eq:ABS}
\sqrt{\Omega} = s \sqrt{2 E_J} \sin \frac{\varphi}{2}.
\end{equation}
 Here, $E_J$ is the Josephson energy associated with the  lowest ABS: $E_J=E_J^*$ in the single-channel case, $E_J^*>E_J$ in general, and $s=\pm1$ characterizes the superposition between right/left leads. Equation~\eqref{eq:ABS} with $s={\rm sign}(\sin\frac{\varphi}{2})$ reproduces the ABS dispersion in a short junction in the tunnel limit~\cite{Catelani2012}.
With the environment, the above relation is modified to a singular-value equation for a wave function $|\Phi\rangle$ in the environmental degrees of freedom, that involves {\it square-roots} of Hamiltonian-like operators, 
\begin{equation}
\label{answer-ham}
\left(\sqrt{\Omega + H} - s \sqrt{2 E_J} \sin \frac{\hat{\varphi}}{2}\right) |\Phi\rangle =0,
\end{equation}
where $H=H_{\rm e}-E_g^{(e)}$ with $E_g^{(e)}$ the ground state energy in the even parity sector.

The path integral approach is also non-trivial, bearing a similarity with the Green function treatment of a frozen disorder~\cite{AGD}. The key quantity is a propagator $G(\tau, \tau')$ defined in a rather standard way
\begin{equation}
G(\tau,\tau') = G_0(\tau-\tau') 
+ \int d \tau_1 G_0(\tau-\tau_1) A(\tau_1) G(\tau_1,\tau').
\end{equation}
Here $A(\tau) \equiv s \sqrt{2 E_J} \sin \frac{\varphi(\tau)}{2}$ plays the role of the disorder, $G_0(\tau) \equiv \Theta(\tau)/\sqrt{\pi \tau}$ is the bare propagator arising from the reduction of quasiparticle continuum states. The disorder averaging should be done with the weight $e^{-{\cal S}}$,
that is, with respect to the even parity ground state. The averaged propagator is uniform, its Fourier component reads
\begin{equation}
\bar {G}(\omega) = \left( \sqrt{i \omega} - \langle A\rangle -\Sigma(\omega)\right)^{-1},
\end{equation}
the self-energy $\Sigma(\omega)$ being a sum of diagrams involving the correlators of $A(\tau)$ starting from the second order. Finally, the binding energy is found from
\begin{equation}
\label{answer-path}
\sqrt{\Omega} = \langle A\rangle +\Sigma(-i \Omega).
\end{equation}
Equations \eqref{answer-ham} and \eqref{answer-path} demonstrate an involved structure of the resulting theory that is distinct from straightforward Hamiltonian or path-integral approaches. Nevertheless, we manage to get to experimentally verifiable  predictions by using perturbation theory and renormalizations.

Let us start with the case of small ohmic impedance, $\alpha \ll 1$. For a concrete model, we add a capacitance and an inductance  in parallel to the resistor $R$, $Z(\omega) = 1/(-i\omega C + 1/R+i/\omega L)$. This cuts the ohmic response both at high and low frequency, $\omega_H = 1/RC$ and $\omega_L = R/L$, respectively.  The inductance  providing the low cut-off is required in order to phase bias the junction, $E_J e^2 L \ll 1$. 
(The opposite regime may be addressed as in \cite{Hekking1997} for the even case.)
 We concentrate on the single-channel case of quasiparticle poisoning: the odd ground state energy $E^{(o)}_{g}$ does not depend on phase without fluctuations. We aim to compute the phase-dependent correction $\delta E^{(o)}_{g}\hspace{-2pt}(\varphi)$ proportional to the fluctuations, which defines the supercurrent in the odd state.

\begin{figure}
\includegraphics[width=\columnwidth]{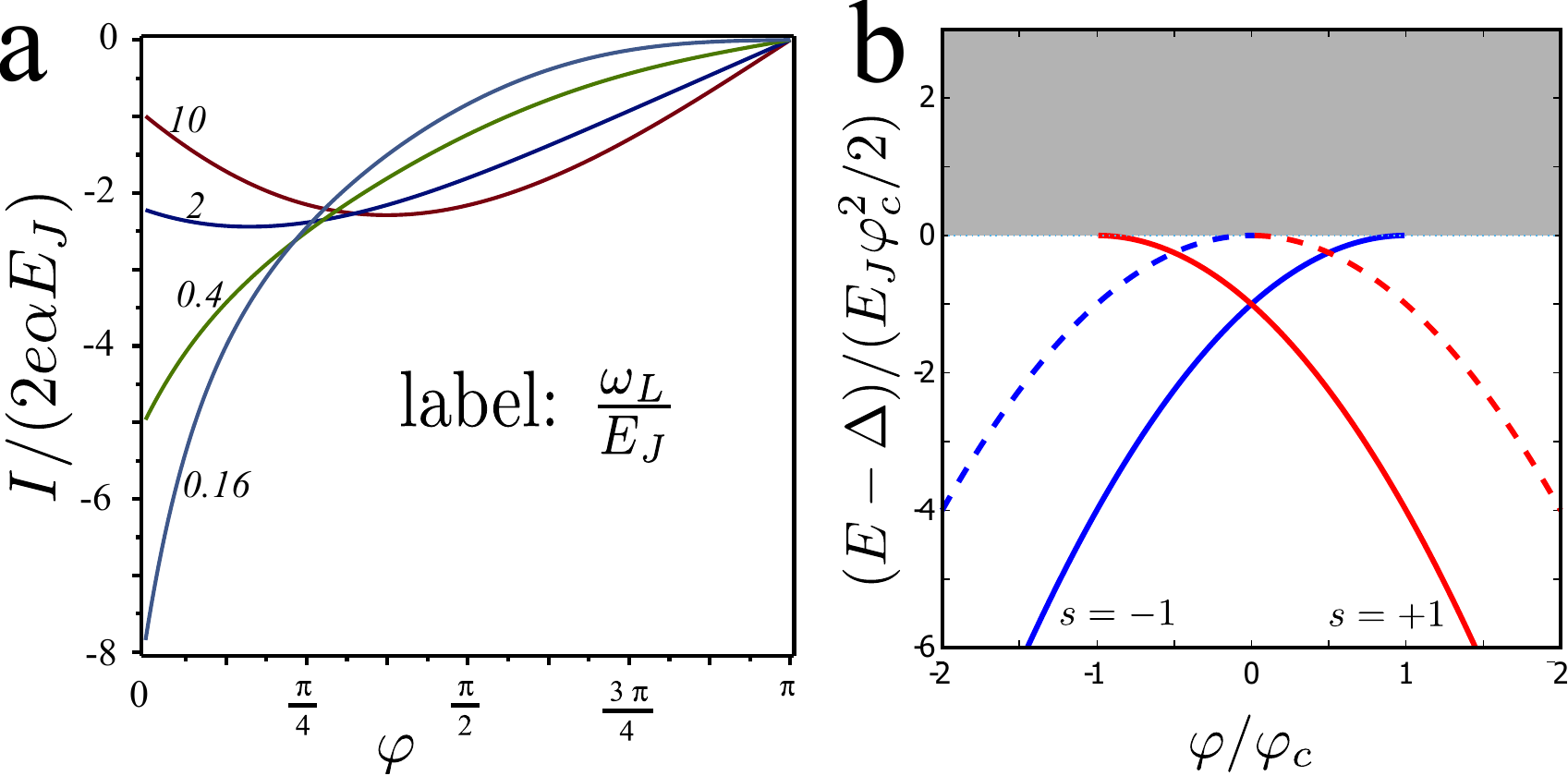}
\caption{\label{fig:small}
a. The odd-parity supercurrent at small impedance. The curve labels are $\omega_L/E_J$, we set $\ln(\omega_H/\omega_L) =5$.
b. Bound states near zero phase for $s=\pm 1$.  Here $\varphi_c= \pi\alpha\sqrt{E_J/\omega_L}\ll1$. Dashed curves: no interaction. }
\end{figure}

A simple ad hoc estimation would be $ \delta E^{(o)}_{g} \simeq \alpha E_J \cos \varphi$. While this may be a correct scale, the answer is more involved and interesting,  see Fig.~\ref{fig:small}a. We note an extra dimensionless parameter $\omega_L/E_J$ that can be large or small provided $\alpha \ll 1$.  
We see that the current  in the phase interval $\varphi\in[0,\pi]$ is {\it negative}: the minimum odd-parity Josephson junction energy corresponds to $\varphi=\pi$ rather than $\varphi=0$.
Let us note that this $\pi$-junction behaviour has a completely different origin than the one induced by magnetic correlations in the ground state of a superconductor-quantum dot-superconductor junction~\cite{Bulaevskii1977} or the one due to the continuum contribution that is left in the presence of poisoning when weak interactions and a finite length of the junction are taken into account~\cite{Kurilovich2021}.
At $\omega_L \gg E_J$, 
\begin{equation}
\label{eq:curr1}
\frac{I(\varphi)}{2e} = - \frac{\alpha E_J}{2}{\ln} \left(\frac{\omega_H}{\omega_L}\right){\sin\varphi}.
\end{equation}
The most interesting feature present for arbitrary ratios $\omega_L/E_J$ is the current jump at $\varphi=0$, its half-value being
\begin{equation}
\label{eq:curr2}
\frac{I_{{\rm hj}}}{2e} = - \pi \alpha E_J \sqrt{\frac{E_J}{\omega_L}}.
\end{equation}
At $\omega_L \ll E_J$, the supercurrent is concentrated at small $\varphi \simeq \sqrt{\omega_L/E_J}$ and reads
\begin{align}
\label{eq:curr3}
I(\varphi) = - |I_{\rm hj}| f(\varphi/\sqrt{2\omega_L/E_J})
\end{align}
with $f(0)=1$ and $f(x) \to \sqrt{2}/\pi x$  at $x \to \infty$. The full expression for the monotonous function $f(x)$ is given in~\cite{Note1}.

The current jump is associated with the fact that
the perturbation theory formally ceases to hold at small $\varphi$. However, the answer beyond perturbations is really simple  and shown in Fig. \ref{fig:small}b:  namely the binding energy is shifted such that the bound state reaches the continuum edge not at $\varphi=0$, but at $\varphi=-s\varphi_c$ with $\varphi_c \equiv (|I_{\rm hj}|/2e)/E_J$, i.e., the binding energy is given by 
\begin{equation}
\label{eq:smallphistructure}
\sqrt{\Omega} = \sqrt{E_J/2} \left( s \varphi + \varphi_c\right).
\end{equation}
The shifts being opposite for $s = \pm 1$, this implies 
the  presence of bound states for {\it both} superpositions in an interval $|\varphi|<\varphi_{c}$: this fact will become crucial for further analysis.

Let us turn to the case of an arbitrary impedance, $\alpha \simeq 1$, under conditions of {\it phase bias}. In this case, the low cut-off frequency  is such that $\omega_L\gg E_J$ and does not change upon renormalization of $E_J,E^*_J$. {The} renormalization is thus finite at any $\alpha$: this implies that, as discussed in the even parity sector~\cite{Hekking1997}, no Schmid transition {occurs under  phase bias}.
{While $E_J=E_J^*$ in the single-channel case, they renomalize differently.}
The renormalization can be computed using the relation $\langle e^{i \beta \varphi}\rangle = e^{i \beta \langle \varphi\rangle} e^{-\beta^2 \langle\hspace{-2pt}\langle \varphi^2\rangle\hspace{-2pt}\rangle/2}$, where $\langle\hspace{-2pt}\langle\varphi^2\rangle\hspace{-2pt}\rangle=\langle\varphi^2\rangle-\langle\varphi\rangle^2$, valid for Gaussian fluctuations of the phase. At even parity, 
\begin{equation}
\label{eq:evenrenormalization}
\tilde{E}^*_J = E^*_J e^{-\langle\hspace{-2pt}\langle {\varphi}^2\rangle\hspace{-2pt}\rangle /2} \simeq E^*_J \left( {\omega_L}/{\omega_H}\right)^\alpha.
\end{equation}
Here and further on the \lq tilde\rq\ refers to renormalized quantities.

\begin{figure}
\includegraphics[width=\columnwidth]{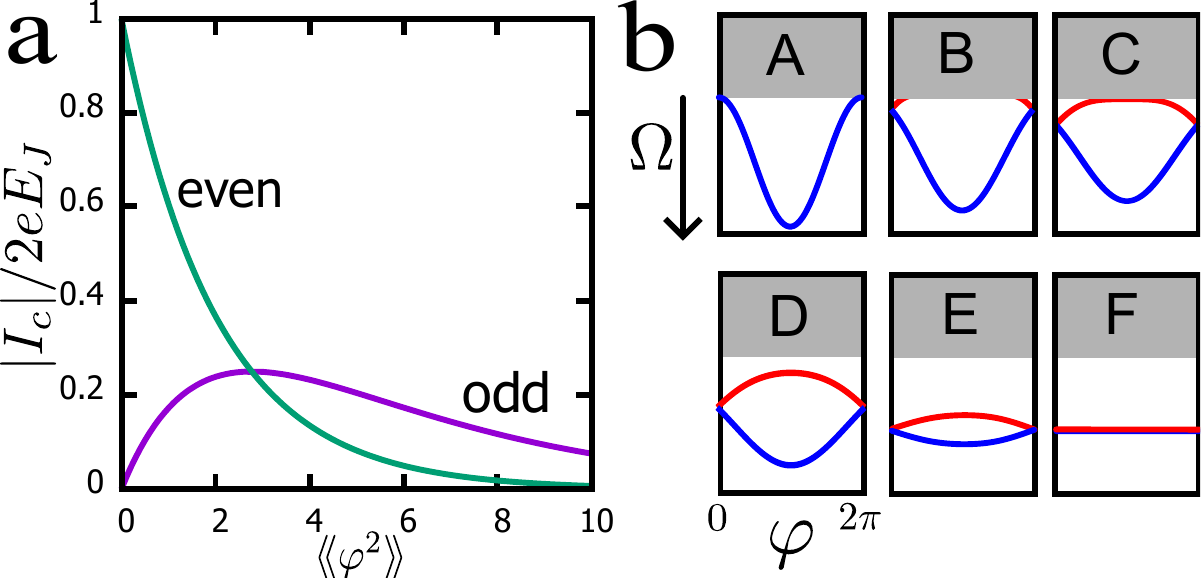}
\caption{\label{fig:fig3}
a. Critical currents at even and odd parity versus $\langle\hspace{-2pt}\langle \varphi^2 \rangle\hspace{-2pt}\rangle $, Eq.~\eqref{eq:current-rem}. 
The odd parity current  
dominates at $\langle\hspace{-2pt}\langle \varphi^2 \rangle\hspace{-2pt}\rangle > 4\ln2\approx 2.8$. 
b. The bound regimes in the odd parity Josephson junction. A: only one superposition gives rise to a bound state ($\alpha =0$). B: two bound states in a finite phase interval (cf.~Fig.~\ref{fig:small}b). C:  separatrix between B and D. D: two $4\pi$-periodic bound states are present at all phases. 
E: the splitting of the two bound states is much smaller than their average phase-independent energy. F: The two states $s=\pm1$ with phase-independent energy are degenerate.
}
\end{figure}

To understand the renormalization at odd parity, we keep  terms up to the second order in the self-consistency equation (\ref{answer-path}),
\begin{equation}
\label{eq:intermediate}
\sqrt{\Omega} = \langle A \rangle + \Sigma^{(2)}(-i \Omega).
\end{equation}
The average $A$ is phase-dependent and strongly suppressed,
\begin{equation}
\label{eq:A-renormalization}
\langle A \rangle = s\sqrt{2 \tilde{E}_J} \sin\frac \varphi 2\quad{\rm with}\quad \frac{\tilde{E}_J}{E_J} = e^{- \frac{\langle\hspace{-2pt}\langle \varphi^2 \rangle\hspace{-2pt}\rangle}{4}} \simeq \left( \frac{\omega_L}{\omega_H}\right)^{\frac \alpha 2}.
\end{equation}
This suppression is two times weaker than for even parity. 
The superconducting current in the odd state at $\alpha <2$ reads
\begin{equation}
\label{eq:current-rem}
\frac{I(\varphi)}{2e} = (\tilde{E}^*_J - \tilde{E}_J) \sin\varphi = E_J \left(  e^{-\frac{\langle\hspace{-2pt}\langle\varphi^2\rangle\hspace{-2pt}\rangle}{2}} - e^{-\frac{\langle\hspace{-2pt}\langle \varphi^2\rangle\hspace{-2pt}\rangle}{4}} \right) \sin\varphi
\end{equation}
and is bigger than that at even parity at sufficiently large phase fluctuations, see Fig.~\ref{fig:fig3}a.   

However,  as far as the bound state spectrum is concerned, the second-order term $\Sigma^{(2)}(-i \Omega)$ can become important since it has a phase-independent part.  This leads to a variety of {\it bound} regimes A-F listed in Fig. \ref{fig:fig3}b. For estimates, we concentrate on 
the phase-independent terms in $\Sigma^{(2)}$
and, since $\Omega \ll \omega_L$, disregard the $\Omega$ dependence. This yields
\begin{equation}
\Sigma^{(2)} = {E_J} \int_0^{\infty} \frac{d \tau}{\sqrt{\pi \tau}} \langle\hspace{-2pt}\langle e^{i \varphi(0)/2} e^{-i \varphi(\tau)/2}\rangle\hspace{-2pt}\rangle.
\end{equation}
The integrand at $\omega_H^{-1} \ll \tau \ll \omega_L^{-1}$ is proportional to $1/\tau^{1/2}(\omega_H \tau)^{\alpha/2}$.  As a consequence, the integral converges at the lower cut-off if $\alpha <1$ and at the upper cut-off if $\alpha >1$. The estimations for $\Sigma^{(2)}$ thus read:
\begin{align}
\label{eq:sigmaest1}
\Sigma^{(2)} \simeq \left\{\begin{array}{ll}
{\tilde{E}_J}/{\sqrt{\omega_L}},& \alpha<1,
\\
{E_J}/{\sqrt{\omega_H}},& \alpha>1.
\end{array}\right.
\end{align}
Comparing $\langle A \rangle$ at $\varphi=\pi$ and $\Sigma^{(2)}$, we observe that the latter dominates for $\alpha > 2[1+ \ln(\omega_L/E_J)/\ln(\omega_H/\omega_L)]\equiv \alpha_c >2$. This point (C) separates two different regimes. Now we can summarize the results.

At $\alpha <\alpha_c$, $\Sigma^{(2)}$ can be neglected in zeroth approximation. The superconducting current is given by
Eq.~\eqref{eq:current-rem}. Starting from the case without fluctuations (regime A), we find that the addition of small phase-independent terms in Eq.~\eqref{eq:intermediate} when fluctuations are weak leads to the coexistence of  two bound states corresponding to the two superpositions $s=\pm1$ in a small interval of phases around $\varphi=0$ (regime B). At $\alpha>1$, this interval grows with increasing $\alpha$ until an important transition (regime C) takes place at $\alpha_c$: two bound states are present at any phase. For $\alpha>\alpha_c$, both bound states are separated from the continuum by a gap  ({regime D}). Thus the odd parity state becomes stable upon an adiabatic sweep of the phase. The bound energies are given by 
\begin{equation}
\Omega = \left(s \sqrt{2 \tilde{E}_J} \sin \frac{\varphi}{ 2} +  \Sigma^{(2)}\right)^2.
\end{equation}
The resulting superconducting current at a given $s$ thus becomes $4\pi$-periodic, a phenomenon similar to that signifying the presence of Majorana modes \cite{Mayorana4pi}. The $2\pi$-periodicity is restored upon relaxation to the lowest energy state within the odd sector.

At $\alpha$ slightly (by $\simeq 1/\ln(\omega_H/\omega_L) \ll 1$) exceeding $\alpha_c$, the binding energy $\Omega \simeq E_J^2/\omega_H \,\gg\, \tilde E_J$ hardly depends on the phase and $\alpha$ (regime E). The remaining phase dependence results in a strongly suppressed $4\pi$-periodic supercurrent 
\begin{equation}
\label{eq:almostdegenerate}
 \frac{I(\varphi)}{2e} \, \simeq\,  s \tilde E_J^{\rm eff} \cos \frac{\varphi}{2}; \; \tilde E_J^{\rm eff} \simeq \sqrt{\tilde{E}_J \Omega} \simeq E_J \sqrt{\frac{\tilde{E}_J}{\omega_H}}.
\end{equation}
Despite being suppressed, this supercurrent parametrically exceeds the one at even parity.

\begin{figure}
\includegraphics[width=\columnwidth]{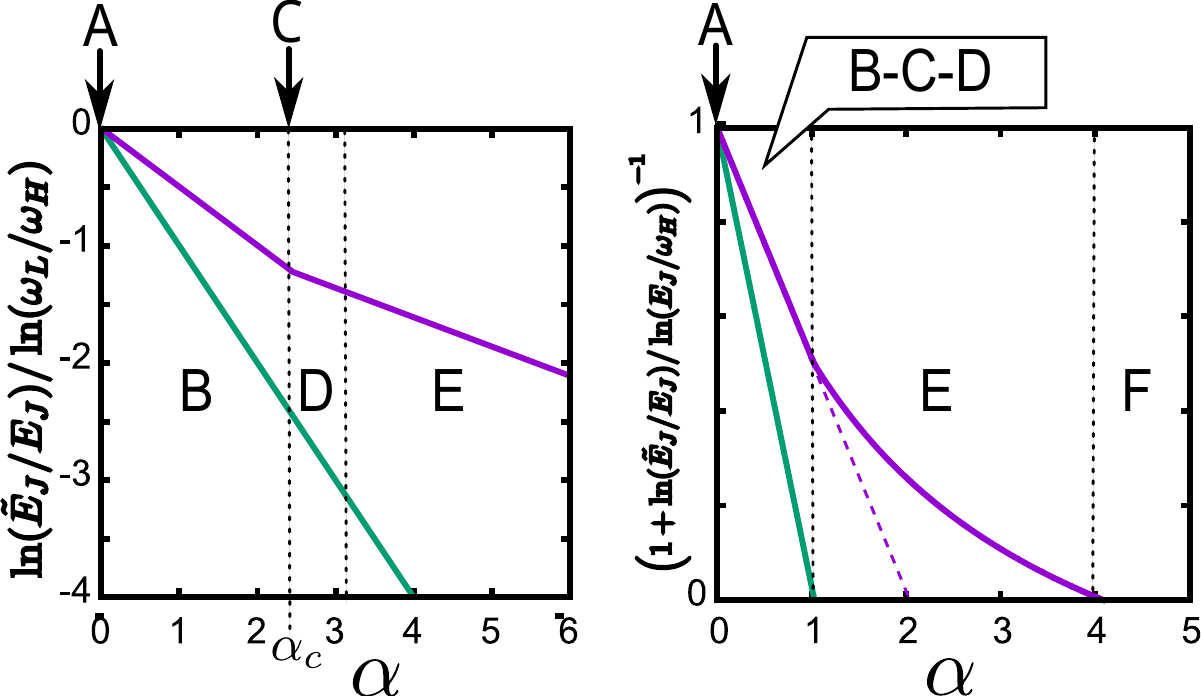}
\caption{\label{fig:fig4}
Renormalized Josephson energies  $\tilde{E}_J^*$ (green) at even and $\tilde{E}_J$ (violet) at odd parity. Vertical dotted lines separate the bound regimes at odd parity indicated by capital letters.  Left: phase bias, cf.~Eqs.~\eqref{eq:evenrenormalization}, \eqref{eq:A-renormalization}, and \eqref{eq:almostdegenerate}; $\tilde{E}_J$ never vanishes.  The separating regime $C$ occurs at $\alpha=\alpha_c$. We plot $\tilde{E}_J^{\rm eff}$ instead of $\tilde{E}_J$ at $\alpha>\alpha_c$.  Right: current bias, cf.~Eqs.~\eqref{eq:Ibias-even}, \eqref{eq:Ibias-odd1}, and \eqref{eq:Ibias-odd2}; the curves illustrate the suppression of $\tilde E_J$ as $\alpha$ increases, the Schmid transition where $\tilde E_J$ vanishes is at $\alpha=1$ for even parity and at $\alpha=4$ for odd parity. The renormalization law at odd parity changes at $\alpha=1$. Note the different vertical scales in the left and right plot. }
\end{figure}

Let us  now turn to the case of an arbitrary impedance at {\it current bias}, see Fig.~\ref{fig:FirstFigure}c. In contrast with the phase bias situation, there is no built-in low energy cut-off  $\omega_L$:  the renormalization has to be cut-off self-consistently by the renormalized Josephson energy.

Let us first reproduce the Schmid transition at even parity. The renormalized $\tilde{E}^*_J$ is given by the same Eq.~\eqref{eq:evenrenormalization},
yet $\omega_L$ there has to be estimated as $\tilde{E}^*_J$. With this,
\begin{equation}
\label{eq:Ibias-even}
\frac{\tilde{E}^*_J}{E^*_J} = \left( \frac{E^*_J}{\omega_H}\right)^\frac{\alpha}{1-\alpha},
\end{equation}
such that ${\tilde{E}^*_J}$ vanishes at the Schmid transition, $\alpha=1$.  

Let us  next turn to the odd parity sector. To start with, let us concentrate on the interval $\alpha<1$. In this case, the lower cut-off can be unambiguously identified as $\tilde{E}_J$. Applying Eq.~\eqref{eq:A-renormalization}, we  thus obtain
\begin{equation}
\label{eq:Ibias-odd1}
\frac{\tilde{E}_J}{E_J} = \left( \frac{E_J}{\omega_H}\right)^\frac{\alpha}{2-\alpha}.
\end{equation}
The estimation of $\Sigma^{(2)}$ with the help of Eq.~\eqref{eq:sigmaest1} gives $\Sigma^{(2)} \simeq \sqrt{\tilde{E}_J}$. In contrast with the phase-biased case, the first- and second-order contributions are of the same order of magnitude, as well as  all higher orders. So in the current bias case, the accuracy of the method does not allow to predict the phase dependence of the energy, nor if bound states persist for both values of $s$ {(regimes B-C-D)}. 

However, we still may notice and use the difference in  the renormalizations of phase-dependent and phase-independent parts of $\sqrt{\Omega}$ depending on the value of $\alpha$. This becomes important at $\alpha >1$ where, in accordance with Eq.~\eqref{eq:sigmaest1}, the  self-energy $\Sigma^{(2)}$ does not depend on the  low-energy cut-off anymore and saturates at the value $\simeq E_J/\sqrt{\omega_H}$. 
As to the phase-dependent part, it will further decrease with increasing $\alpha$. This brings us to {regime E}: the almost degenerate bound state  associated with the supercurrent described by Eq.~\eqref{eq:almostdegenerate}. In this case,  the renormalization of $E_J$ is cut off by $\tilde E_J^{\rm eff}$ of Eq.~\eqref{eq:almostdegenerate}, rather than $\tilde{E}_J$. This yields
\begin{equation}
\label{eq:Ibias-odd2}
 \frac{\tilde{E}_J}{E_J} = \left(\frac{E_J}{\omega_H}\right)^\frac{3\alpha}{4-\alpha}
 ; \;
 \frac{\tilde{E}_J^{\rm eff}}{E_J} = \left(\frac{E_J}{\omega_H}\right)^{\frac{\alpha+2}{4-\alpha}}
 .
\end{equation} 
Therefore $\tilde{E}_J$,$\tilde{E}^{\rm eff}_J$ vanish at $\alpha=4$. This is the new Schmid transition point for half of the Cooper pair charge, indeed corresponding to $4\pi$-periodicity in phase of the supercurrent.
At $\alpha>4$, the phase-independent bound state is completely degenerate with respect to $s$ ({regime F}). Recalling the  quasiparticle spin, we  thus predict the realization of 4-fold degeneracy for the trapped quasiparticle.

In conclusion, we have formulated the Josephson quantum mechanics for a junction in the odd parity state. The non-trivial structure of the theory is encapsulated in Eqs.~\eqref{answer-ham} and \eqref{answer-path}. 
We concentrated on the single-channel case and predicted the lifting of  the supercurrent quench due to quasiparticle poisoning at small $\alpha$. The residual supercurrent is given by Eqs.~\eqref{eq:curr1}-\eqref{eq:curr3}. 
Furthermore, we have addressed the case of arbitrary impedance both at phase and current bias. The supercurrent at odd parity is less suppressed  by quantum fluctuations and  may dominate  over the one at even parity. The presence of various bound regimes complicates the renormalization. For current bias, we predict a Schmid transition at $\alpha =4$ and four-fold degenerate bound states at higher impedances.

\begin{acknowledgments} 
YVN acknowledges support from the Université Grenoble Alpes for an extended stay in Grenoble during which most of the presented work was performed. MH and JSM acknowledge funding from the Plan France 2030 through the project NISQ2LSQ ANR-22-PETQ-0006.
\end{acknowledgments}

\bibliography{josodd-bib}

\end{document}


\title{Josephson quantum mechanics at odd parity}

\author{Manuel~Houzet, Julia~S.~Meyer}
\affiliation{Univ.~Grenoble Alpes, CEA, Grenoble INP, IRIG, Pheliqs, F-38000 Grenoble, France}
\author{Yuli V. Nazarov}
\affiliation{Kavli Institute of Nanoscience, Delft University of Technology, 2628 CJ Delft, The Netherlands}


\maketitle

\let\oldvec\vec
\renewcommand{\vec}[1]{\ensuremath{\boldsymbol{#1}}}



	
\section*{Supplemental Material}
In this Supplemental Material, we present the details of the derivation of our results.

\section{Odd parity ground state: no quantum fluctuations}

Here we recall the derivation of Andreev bound states in the tunneling limit closely following Appendix D of Ref.~\cite{Catelani2012}. For the moment, we disregard the fluctuations of the phase treating it as a number.

We describe a Josephson junction with $N_{\rm ch}$ tunneling channels by the Hamiltonian
\begin{equation}
H=\sum_{km\sigma}\varepsilon_{k}\alpha_{km\sigma}^\dagger \alpha_{km\sigma}+\sum_{km\sigma}\varepsilon_{k}\gamma_{km\sigma}^\dagger \gamma_{km\sigma} +H_T.
\end{equation}
Here $\alpha_{km\sigma}$ ($\gamma_{km\sigma}$) is a fermionic annihilation operator of a Bogoliubov quasiparticle in the left (right) lead, with orbital label $k$, spin $\sigma$, and excitation energy $\varepsilon_{k}$, which participates in the $m^{\rm th}$ tunneling channel of the junction ($1\leq m\leq N_{\rm ch}$, where $N_{\rm ch}$ is the number of channels). 
The quasiparticle energies are  $\varepsilon_k=\sqrt{\xi^2_k+\Delta^2}$, where $\Delta$ is the superconducting gap and $\xi_k$ is the electron energy of the state $k$ measured from the Fermi level in the absence of superconductivity.

The tunneling is diagonal in channels, the corresponding term reads ($L$ being the infinite normalization length of the channel) 
\begin{equation}
\label{eq:HT}
H_T=e^{i\varphi/2}\sum_{kk'm\sigma}\frac{t_m}{L} a_{km\sigma}^\dagger c_{k'm\sigma}+{\rm H.c.}.
\end{equation}
It is expressed in terms of the annihilation operator of an electron in the right lead, $a_{km\sigma}=u_k \alpha_{km\sigma}+\sigma v_k\alpha^\dagger_{km-\sigma}$, with coherence factors $u_k,v_k=\sqrt{(1\pm\xi_k/\varepsilon_k)/2}$, and the annihilation operator in the left lead $c_{km\sigma}$ expressed in terms of $\gamma_{km\sigma}$ in a similar way. The tunnel matrix element $t_m$  is real by virtue of time reversibility and defines the transmission coefficient $T_m$ of the corresponding channel, $T_m= (2 \pi \nu t_m)^2\ll 1$, $\nu$ being the density of states per spin and channel.

Following Appendix D of Ref.~\cite{Catelani2012}, we note that an effective low-energy description of the odd parity sector involves the states with energies close to $\Delta$, so that $u_k\approx v_k\approx 1/\sqrt{2}$. With this, the tunneling Hamiltonian becomes
\begin{equation}
\label{eq:HTred}
H_T= \sum_{m,\sigma} i\frac{t_m}{L}\sin\frac{\varphi}{2}\sum_{kk'}\left(\alpha^\dagger_{km\sigma}\gamma_{k'm\sigma}-\gamma^\dagger_{km\sigma}\alpha_{k'm\sigma}\right).
\end{equation}
It involves only the quasiparticle transfers and no terms creating/annihilating a pair of quasiparticles. 

To get further insight into the specific structure of the tunneling Hamiltonian, let us introduce two linear combinations of the left and right quasiparticle operators,
\begin{equation}
\beta_{kms\sigma}=(\alpha_{km\sigma} - i s\gamma_{km\sigma})/\sqrt{2}
\end{equation}
with the index $s=\pm 1$ that we call {\it superposition index},  or chirality. The operators create a quasiparticle in an equal-weight superposition of the states in the left and right lead. 
 
Switching to these new operators, we observe that $H_T$ conserves the superposition index,
and the full Hamiltonian describing the quasiparticles close to the gap edge reads
\begin{equation}
\label{eq:Ho}
H_{{\rm full}}=\sum_{kms\sigma}\varepsilon_k\beta_{kms\sigma}^\dagger\beta_{kms\sigma}- \frac{t_m}{L}\sin\frac{\varphi}{2}\sum_{kk' ms\sigma} s \beta_{kms\sigma}^\dagger\beta_{k'ms\sigma}
\end{equation}
with $\varepsilon_k\approx\Delta+\xi_k^2/2\Delta$. 

Let us consider a single quasiparticle with given $\sigma$ and $s$ in a given channel $m$.
The Hamiltonian in the space of the possible states $k$ of this quasiparticle reads
\begin{equation}
\label{eq:oneparticle}
H_{{\rm qp}} = \Delta + \sum_{k}\frac{\xi_k^2}{2\Delta} |k\rangle \langle k| - s \sin\frac{\varphi}{2} \frac{t_m}{L} \sum_{k,k'}  |k\rangle \langle k'|.
\end{equation}
The eigenstates with $E >\Delta$ correspond to delocalized quasiparticles and are of no interest to us. We concentrate on possible bound states of the quasiparticle with $E=\Delta - \Omega$, $\Omega$ being the positively defined binding energy of the quasiparticle. The Schr\"{o}dinger equation for the wave function $\Psi = \sum_k\psi_k |k\rangle$  reads 
\begin{equation}
0=\left(\Omega + \frac{\xi_k^2}{2\Delta}\right) \psi_k - s \sin\frac{\varphi}{2}\frac{t_m}{L} \sum_{k'} \psi_k' 
\end{equation}
and can be easily solved for $\Phi = \sum_k \psi_k$,
\begin{equation}
\Phi =  s \sin\frac{\varphi}{2}\frac{t_m }{L} \sum_k \frac{\Phi}{\Omega + {\xi_k^2}/{2\Delta}} .
\end{equation}
We replace the sum over $k$ by integration over 
energies, $\sum_k \to L \nu \int d \xi$. The resulting integral over $\xi$ converges at the energies of the order of $\Omega$. This gives 
\begin{equation}
\sqrt{\Omega} = \sqrt{2 E^{(m)}_J} s \sin\frac{\varphi}{2} ,
\end{equation}
where $E^{(m)}_J \equiv \Delta (\pi \nu t_m)^2 \equiv \Delta T_m/4 $ is the contribution of the level $m$ to the total Josephson energy $E^*_J$ in the even ground state, 
$E^*_J = \sum_m E^{(m)}_J$. 
For simplicity, we refer to the contribution of the lowest level, that is, the level with the largest $t_m$, just as $E_J$. 
Therefore, $E^*_J \geq E_J$  in general and $E^*_J = E_J$ for a single-channel situation where the contribution of other channels can be neglected.

\section{Odd parity ground state: Hamiltonian approach}
Now let us take into account quantum fluctuations of $\varphi$. To this end, we promote the phase to an operator. Let us consider first the even parity state. In this case, the Hamiltonian comprises the Hamiltonians of the environment and the junction,
\begin{equation}
H_{{\rm e}} = H_{{\rm env}} - E^*_J \cos \hat \varphi.
\end{equation}
We use a bosonic description of the linear environment,
\begin{equation}
\label{eq:env}
H_{\rm env}=\sum_q\omega_q b_q^\dagger b_q, \qquad \hat \varphi=\varphi+\sum_q\lambda_q(b_q+b^\dagger_q).
\end{equation}
Here $b_q$ is a bosonic annihilation operator of an excitation in mode $q$, with energy $\omega_q$.
The operator of the phase $\hat \varphi$ is composed of the constant phase bias part $\varphi$ and the fluctuating part that is a linear superposition of the bosonic creation/annihilation operators. The coefficients $\lambda_q$ in this superposition are chosen to represent the dissipative part of the environment impedance seen  by the junction, 
\begin{equation}
{\rm Re} Z(\omega)=\frac{\pi\omega}{4e^2}\sum_q\lambda_q^2\delta(\omega-\omega_q), 
\end{equation}
at frequency $\omega>0$.

It is convenient for us to use this Hamiltonian with the energy counted from its ground state $|0\rangle$, and define
\begin{equation}
H = H_{{\rm e}} - E^{(e)}_g, \; \; \; H|0\rangle = 0.
\end{equation}

To obtain the odd parity state, we add a quasiparticle of chirality $s$ to the lowest Andreev state. With this, the total Hamiltonian $H_{{\rm odd}}$ is obtained by combining the Hamiltonian of Eq.~\eqref{eq:oneparticle} and $H$, the binding energy $\Omega$ being an eigenvalue of this Hamiltoninan,
\begin{eqnarray}
H_{{\rm odd}} &=& H + \sum_{k}\frac{\xi_k^2}{2 \Delta} |k\rangle \langle k| - s \sin \frac{\hat{\varphi}}{2}\frac{t_m}{L} \sum_{k,k'}  |k\rangle \langle k'|,
\\
0&=& (\Omega + H_{{\rm odd}}) |\Psi\rangle.
\end{eqnarray}
With all the energy shifts we made, the ground state energy at odd parity is given by
\begin{equation}
E_g^{(o)} = E^{(e)}_g +\Delta - \Omega. 
\end{equation}

The wave function $|\Psi\rangle$ is in the combined space of $k$ and environmental degrees of freedom. Further derivation essentially repeats the steps done in the previous Section. We substitute $|\Psi\rangle = \sum_k|\psi_k\rangle |k\rangle$, $|\psi_k\rangle$ being a wave function in the space of environmental degrees of freedom, and obtain a closed equation for
$|\Phi\rangle \equiv \sum_k|\psi_k\rangle$,
\begin{equation}
\label{eq:mainHamiltonian}
(\sqrt{\Omega + H} - \hat{A})|\Phi\rangle; \quad  \quad \hat{A} \equiv s\sqrt{2 E_J} \sin \frac{\hat{\varphi}}{2}.
\end{equation}
While Eq.~\eqref{eq:mainHamiltonian} is an eigenvalue problem for an operator, it is more complicated than a standard Schr\"{o}dinger equation that can be readily solved by the diagonalization of the operator. For the case in hand, the diagonalization is not enough: as a result of it, one gets a set of eigenvalues that parametrically depend on $\Omega$. 
Since $H$ is a non-negatively defined operator, and $A$ is an operator restricted by $\pm\sqrt{2 E_J}$, at sufficiently big values of $\Omega$, $\Omega > \sqrt{2 E_J}$ all eigenvalues of the set are positive. Decreasing $\Omega$, we achieve the situation where the lowest eigenvalue of the operator is zero. This value of $\Omega$ thus corresponds to the actual binding energy. It can happen that the lowest eigenvalue remains positive up to $\Omega=0$. In this case, there is no bound state (at these settings of $s$ and phase bias $\varphi$).

\section{Odd parity ground state: Path-integral approach}

The path-integral approach to Josephson quantum mechanics (at even parity) has been developed already in the 1980s (see \cite{Schoen1990} for an extensive early review). It has a clear advantage over any Hamiltonian method since it reduces the description to the relevant variable $\varphi(\tau)$ only. Here we consider only the imaginary-time, zero-temperature version of this path-integral method. Within this method, the averages of a product $\prod_j \varphi(\tau_j)$ over the paths represent the averages of the time-ordered products of Matsubara operators $\hat{\varphi}(\tau) \equiv \exp(-H\tau)\hat{\varphi} \exp(H\tau)$ over the {\it ground} state. The weight of a path $\varphi(\tau)$ is given by $\exp(-{\cal S})$, where the action ${\cal S}(\varphi(\tau))$ comprises a quadratic part representing the linear environment and a part representing the Josephson energy, 
\begin{eqnarray}
{\cal S} = {\cal S}_{\rm{env}} - E^*_J \int d\tau \cos \varphi(\tau), \\
{\cal S}_{\rm{env}} = \sum_{\omega} \frac{ \omega }{ {8} e^2 {\cal Z}(\omega)} |\varphi_\omega|^2.
\end{eqnarray}
Here, ${\cal Z}(\omega)$ is the impedance at imaginary frequency. For the simplest circuit of a resistor $R$ and a capacitor $C$ in parallel, ${\cal Z}^{-1}(\omega)\ = R^{-1}\ {\rm sgn}(\omega) + \omega C$, while the usual impedance $Z(\omega) $ at real frequency is given by $Z^{-1}(\omega)= R^{-1} -i \omega C$ for this circuit.

It is worth noting already in the beginning that our approach to the odd parity state does not involve any new action. Rather, we derive and use an implicit equation for the binding energy $\Omega$ that involves path-integral averages over the {\it even} parity state. This distinguishes the present approach from the standard ones.

We start with Eq.~\eqref{eq:mainHamiltonian}. Let us note that the fact that the operator in it has a singular value implies a divergence of the expectation value of the inverse operator over the even parity ground state,
\begin{equation}
{\rm Ex} \equiv \langle 0| \frac{1}{\sqrt{\Omega +H} - \hat{A}} |0 \rangle = \infty.
\end{equation}
This works if $\langle 0|\Phi\rangle \ne 0$, as we can safely assume. In fact, the actual binding energy is determined by the {\it lowest} value of $\Omega$ at which the expectation value diverges. Indeed, Eq.~\eqref{eq:mainHamiltonian} at  $\Omega$ smaller than the binding energy has singular eigenvalues corresponding to the excitations of the environment on the background of the odd parity ground state.

Let us expand the inverse operator in terms of $\hat{A}$ and concentrate on the second term of the expansion,
\begin{equation}
{\rm Ex}^{(2)} = \langle 0| \frac{1}{\sqrt{\Omega +H}} \hat{A} \frac{1}{\sqrt{\Omega +H}} \hat{A} \frac{1}{\sqrt{\Omega +H}} |0 \rangle.
\end{equation}
We represent the inverse square root entering the expansion in terms of an integral over a time-like variable
\begin{equation}
\frac{1}{\sqrt{\Omega +H}}  =\int_{-\infty}^{\infty} d \tau\; G_0(\tau) e^{-\Omega \tau} e^{- H \tau} ; \qquad G_0(\tau) \equiv \Theta(\tau)/\sqrt{\pi \tau},
\end{equation}
where $\Theta(\tau)$ is the Heaviside step function.
With this, we present the term with the expectation value of the product of Matsubara operators,
\begin{equation}
{\rm Ex}^{(2)} = \int d\tau d\tau_1 d\tau_2 \;e^{-\Omega \tau} G_0(\tau_1) G_0(\tau_2-\tau_1) G_0(\tau - \tau_2) \langle 0|\hat{A}(\tau_1) \hat{A}(\tau_2) |0\rangle.
\end{equation} 
The product of the Matsubara operators is already time-ordered, so its expectation value can be replaced with the average over the paths ($A(\tau) \equiv s \sqrt{2 E_J} \sin( \varphi(\tau)/2)$):
\begin{equation}
\langle 0|\hat{A}(\tau_1) \hat{A}(\tau_2) |0\rangle \to \langle A(\tau_1) A(\tau_2) \rangle.
\end{equation}

\begin{figure}
\includegraphics[width=0.5\columnwidth]{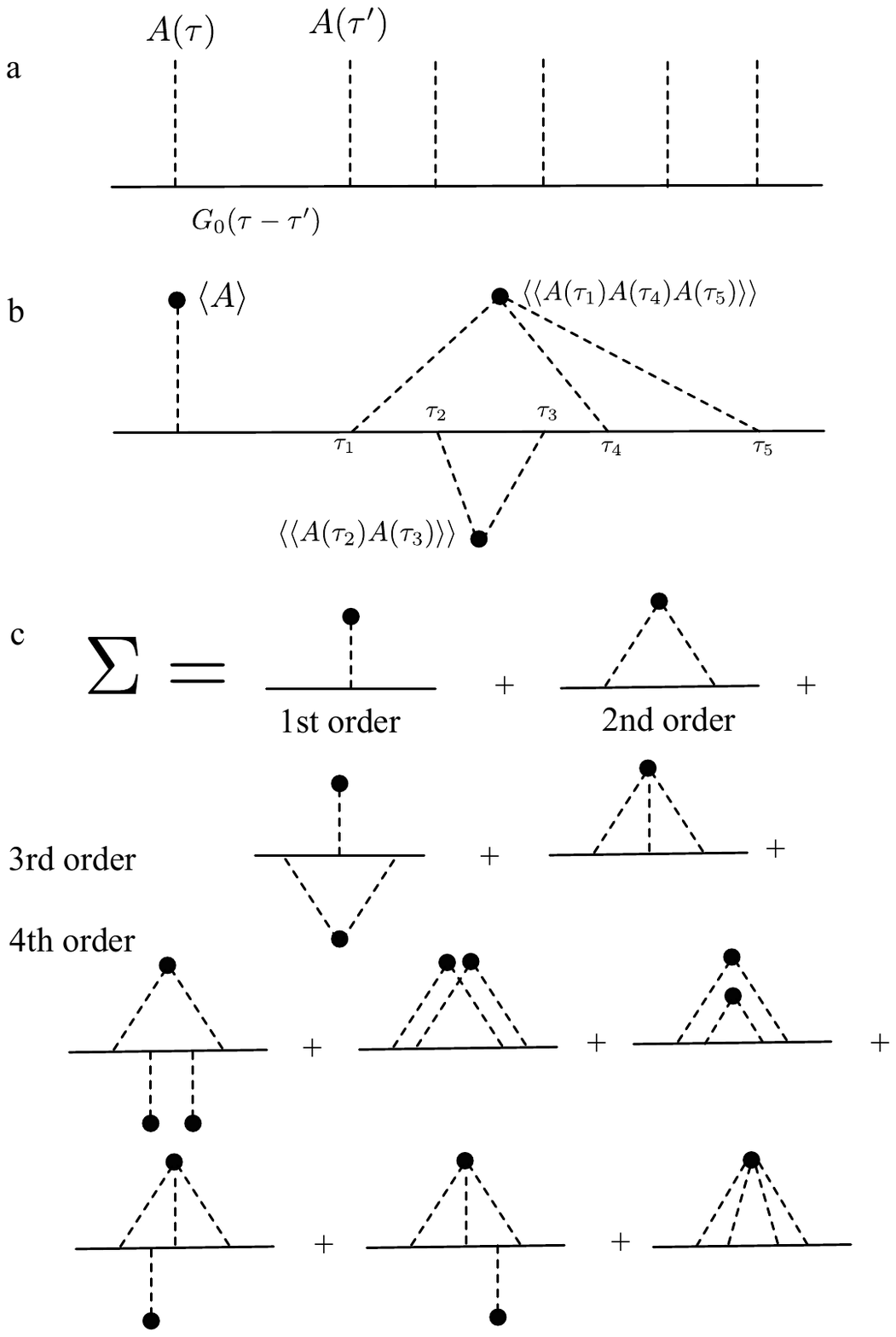}
\caption{\label{fig:selfenergy} The propagator used for the calculation of the binding energy. a. A term in the perturbation expansion of a non-averaged propagator. b. A term in the perturbation expansion of the averaged propagator. c. The diagrams for $\Sigma$ up to 4th order.}
\end{figure}

We do such replacement for all terms of the expansion and sum them up. With this, the expectation value in question is expressed via a propagator $G(\tau,\tau')$ defined by a rather standard relation,
\begin{equation}
G({\tau,\tau'}) = G_0({\tau-\tau'}) + \int d \tau_1 \;G_0({\tau}-\tau_1) A({\tau_1}) G(\tau_1,\tau').
\end{equation}
Let us notice an analogy of the propagator in use and that of a particle in a disordered potential \cite{AGD},
$A(\tau)$ playing the role of the potential. Figure \ref{fig:selfenergy}a gives a diagram presenting a term in the perturbation expansion of $G(\tau,\tau')$.

The average propagator is a function of the time difference only, and
\begin{equation}
\label{eq:integral}
{\rm Ex} = \int d \tau e^{-\Omega t} \bar{G}(\tau) ; \qquad  \bar{G}(\tau) = \langle G(\tau_0+\tau, \tau_0)\rangle.
\end{equation}
The previous reasoning implies that the average propagator grows with increasing time difference, $\bar{G}(\tau) \propto e^{\Omega_{{\rm b}} \tau}$. The integral in \eqref{eq:integral} thus diverges at $\Omega < \Omega_b$ and converges at $\Omega > \Omega_{{\rm b}}$. This identifies $\Omega_{{\rm b}}$ as the actual binding energy.  

The diagram in Fig.~\ref{fig:selfenergy}b represents a term in the expansion of the average propagator. Black dots where the dashed lines come together denote the (higher-order) correlators of $A(\tau)$.  

It would be nice to have a closed expression for $\Omega_{{\rm b}}$, at least a perturbative one. Yet this is beyond our reach. Instead, we use a common wisdom of disorder-averaged propagators \cite{AGD} and introduce a self-energy $\Sigma(\omega)$ such that, in frequency representation, 
\begin{equation}
\bar{G}(\omega) = \frac{1}{ \sqrt{i \omega} - \Sigma(\omega)}.
\end{equation}
The self-energy (in our scheme, its dimension is the square-root of energy) admits a perturbative expression in terms of correlators of $A(\tau)$. The binding energy is defined by the presence of a pole or other singularity at imaginary $\omega$, $ i \omega = \Omega$, this yields
\begin{equation}
\sqrt{\Omega} = \Sigma(-i \Omega),
\end{equation}  
$\Omega_{{\rm b}}$ is thus the root of this equation.
In Fig.~\ref{fig:selfenergy}c we give all diagrams contributing to $\Sigma(\omega)$ up to the fourth order.
This is not the only way to draw the expansion: for instance, one can resum the propagator including all diagramms with $\langle A\rangle $, 
\begin{equation}
G_0 (\omega) \to \bar{G}_0 \equiv \frac{1}{\sqrt{i \omega} - \langle A \rangle},
\end{equation}
so the expansion will start with the second order and would not include black dots with a single line.
We give the explicit expressions for this case up to the fourth order in time representation
\begin{eqnarray}
\Sigma^{(2)}(\tau) &=& \langle\langle A(0) A(\tau) \rangle\rangle \bar{G}_0 (\tau), \\
\Sigma^{(3)}(\tau) &=& \int d\tau_1 \langle\langle A(0) A(\tau_1) A(\tau) \rangle \rangle \bar{G}_0(\tau_1) \bar{G}_0(t-\tau_1), \\
\Sigma^{(4)} (\tau) &=& \int d \tau_1 d \tau_2 \;\bar{G}_0(\tau_1) \bar{G}_0(\tau_2-\tau_1) \bar{G}_0(\tau-\tau_2)  \nonumber\\
&&\times\left( \langle\langle A(0) A(\tau_1) A(\tau_2) A(\tau) \rangle \rangle + \langle\langle A(0) A(\tau) \rangle\rangle \langle\langle A(\tau_1) A(\tau_2) \rangle\rangle + \langle\langle A(0) A(\tau_2) \rangle\rangle \langle\langle A(\tau_1) A(\tau) \rangle\rangle\right).
\end{eqnarray}
Here, the double angle brackets $\langle\langle \dots \rangle\rangle$ denote the cumulant of all variables between the brackets.

\section{Small impedance}
\subsection{Warm-up: single oscillator}
As a warm up, let us consider an 
environment consisting of a single oscillator, such that the phase fluctuation created by it is small as compared to $\pi$. We can proceed with the Hamiltonian method. The Hamiltonian of the oscillator can be written as 
\begin{equation}
H_{osc} = - \frac{E_C}{2} \partial^2_f +\frac{E_L}{2} f^2 = \hbar \omega_0 \left(  \hat{a}^\dagger\hat{a} + \frac{1}{2}\right),
\end{equation}
$f$ being the addition to the phase, $\hat{a}$ being the annihilation operator of the oscillation. Here, $E_L = (4 e^2 L)^{-1}$, $E_C = 4 e^2/C$. The oscillator frequency $\omega_0 = \sqrt{E_C E_L}$ is much smaller than $E_L$, this guarantees the smallness of the fluctuation,
\begin{equation}
\hat{f} = \sqrt{\frac{\omega_0}{2 E_L}} (\hat{a} +\hat{a}^\dagger).
\end{equation}
Here $\omega_0/E_L=\sqrt{L/C}$ is an effective impedance of the oscillator. By virtue of phase bias, $E_L \gg E_J$. As to $E_J$ and $\omega_0$, they can be in an arbitrary relation: this complicates the analysis.

Since the fluctuations are small, we can expand
\begin{eqnarray}
\cos(\hat{\varphi}) \to  \cos(\varphi) - \sin(\varphi) \hat{f} - \cos(\varphi) \frac{\hat{f}^2}{2}, \label{eq:expand1}\\
\sin \frac{\hat{\varphi}}{2} \to \sin \frac{\varphi}{2} +\cos \frac{\varphi}{2} \frac{\hat{f}}{2} -\sin\frac{\varphi}{2} \frac{\hat{f}^2}{8}. \label{eq:expand2}
\end{eqnarray}
The relevant corrections to an energy are the first-order corrections in $\langle\langle f^2 \rangle\rangle$ and eventually $\langle f \rangle$ if present, and the second-order corrections in $\hat f$. 

Let us start with the even parity correction. The first and second order corrections read correspondingly to the two terms in the following equation: 
\begin{equation}
\delta E_g^{(e)} =  E^*_J \cos\varphi \frac{\langle 0|\hat{f}^2|0\rangle}{2} - \frac{|\langle 0| E_J^{*}  \sin\varphi \hat{f}|1\rangle|^2}{\omega_0}
\end{equation}
with
\begin{equation}
 \langle 0|\hat{f}^2|0\rangle = \frac{\omega_0}{2 E_L}; \qquad \langle 0| \hat{f} | 1 \rangle =  \sqrt{\frac{\omega_0}{2 E_L}}; \qquad\langle \hat f \rangle = -( E^*_J / E_L)\sin \varphi,
 \end{equation}
 yielding
 \begin{equation}
\label{eq:sosceven} \delta E_g^{(e)} = E^*_J \frac{ \,\omega_0}{ 4 E_L} \cos\varphi  +  \frac{ E_J^{*2} }{ 2 E_L}\sin^2\varphi.
\end{equation}
The first term is identified as the renormalization of $E_J^*$, while the second is an energy induced in the inductance by the supercurrent in the junction. It causes also a shift in $f$, $\delta f = -( E^*_J / E_L)\sin \varphi $ .
As we see, the relative strength of the corrections reflects the ratio $\omega_0/E_J^*$.

Let us turn to odd parity. The correction $\delta \Omega$ to the unperturbed binding energy $\Omega_0 = \sqrt{2 E_J} \sin \frac{\varphi}{2}$ is computed from
\begin{equation}
\frac{\delta \Omega}{2 \sqrt{\Omega_0}} = \delta A +\left(\frac{\sqrt{2 E_J}\cos\frac{\varphi}{2}}{2}\right)^2 \frac{\langle 0| \hat{f} | 1\rangle^2}{\sqrt{\Omega_0 +\omega_0} - \sqrt{\Omega_0}},
\end{equation}
where the first and second term give the first-order and the second-order correction, respectively.  

The correction to $A$ is given as
\begin{equation}
\delta A = \sqrt{ 2 E_J} \left (\cos \frac{\varphi}{2} \frac{\delta f}{2} - \sin\frac{\varphi}{2} \frac{\hbar \omega_0}{16 E_L}\right).
\end{equation}
With this,
\begin{equation}
\delta \Omega = 
- E_J \sin^2 \frac{\varphi}{2} \frac{\hbar \omega_0}{4  E_L} 
- \sin^2 \varphi \frac{E_J E^*_J}{E_L}
+ \frac{E_J^2 \sin^2 \varphi}{2 E_L} \frac{\sqrt{1 + \omega_0/\Omega_0}+1}{2}.
\end{equation}

Let us note that in the classical limit, $\omega_0 \to 0$, the correction to the ground state energy is reduced to the inductive energy, 
\begin{equation}
\delta E_g^{(o)} = \delta E_g^{(e)} - \delta \Omega = - \frac{(E_J^* - E_J)^2}{2 E_L} \sin^2 \varphi.
\end{equation}
This energy is zero if $E_J^* = E_J$ owing to the poisoning. In further considerations, we concentrate on this case.
There, the phase dependence of $\delta E_g^{(o)}$ eventually defines the supercurrent, while it is a small correction to the supercurrent otherwise.

With this, we reduce the correction to the odd ground state energy 
to the following expression:
\begin{equation}
\label{eq:single-osc-energy}
\delta E_g^{(o)} = \frac{\omega^2_0}{8 E_L} \frac{{\rm cotan}^2\frac{\varphi}{2}}{\left(1+\sqrt{1 + \omega_0/\Omega_0}\right)^2}.
\end{equation}
This formula does not look self-explaining and needs to be elaborated.

First of all, let us note that the corresponding current is finite in the limit $\varphi \to 0$.
In this limit, $\omega_0 \gg \Omega_0 \approx E_J \varphi^2/2 $ at any ratio $\omega_0/E_J$. Expanding in $\varphi$ till the first order, we find ($\varphi >0$)
\begin{equation}
\delta E_g^{(o)}  \approx \frac{\omega_0^2}{ 8 E_L \zeta} \left ( 1-\frac{\varphi}{\sqrt{\zeta}} \right) ; \qquad \zeta \equiv \frac{\omega_0}{2 E_J}.
\end{equation}
So the current jumps at $\varphi=0$, the value of the half-jump is given by 
\begin{equation}
\label{eq:currentjump}
\frac{I_{\rm hj}}{2e} = - \frac{E_J}{2} \sqrt{\frac{E_J}{E_L}} \sqrt{\frac{\omega_0}{2 E_L}} \ll E_J.
\end{equation} 

We note that the current is negative at positive phase. In general, the minimum of $\delta E_g^{(o)}$ is achieved at $\varphi=\pi$. 

Let us address the limiting cases. If $\omega_0 \gg E_J$, the current reduces to
\begin{equation}
\frac{I(\varphi)}{2e} = - \frac{E_J}{8} \frac{\omega_0}{E_L} \sin(\varphi).
\end{equation}
This looks like a renormalization of $E_J$ by the small oscillator's effective impedance ${\omega_0}/{E_L}=\sqrt{L/C}$. The current jump is small and can be neglected at this scale, except for very small $\varphi$.

The opposite limiting case $\omega_0 \ll E_J$ is trickier. 
Here, the current of the order of $I_{\rm hj}$ is concentrated in a narrow interval of phase $ \simeq \sqrt{\zeta}$,
\begin{equation}
\label{eq:narrowcurrent}
I(\varphi) = - |I_{\rm hj}| f(\varphi/2\sqrt{\zeta});\qquad f(x) \equiv \frac{1}{\sqrt{1 +x^2} \left(x + \sqrt{1 +x^2}\right)^2}\enspace{\rm with}\enspace f(0)=1, \enspace f(x) \to 1/(8x^3) \enspace {\rm at} \enspace x \to \infty.   
\end{equation} 

\begin{figure}
\includegraphics[width=0.5\columnwidth]{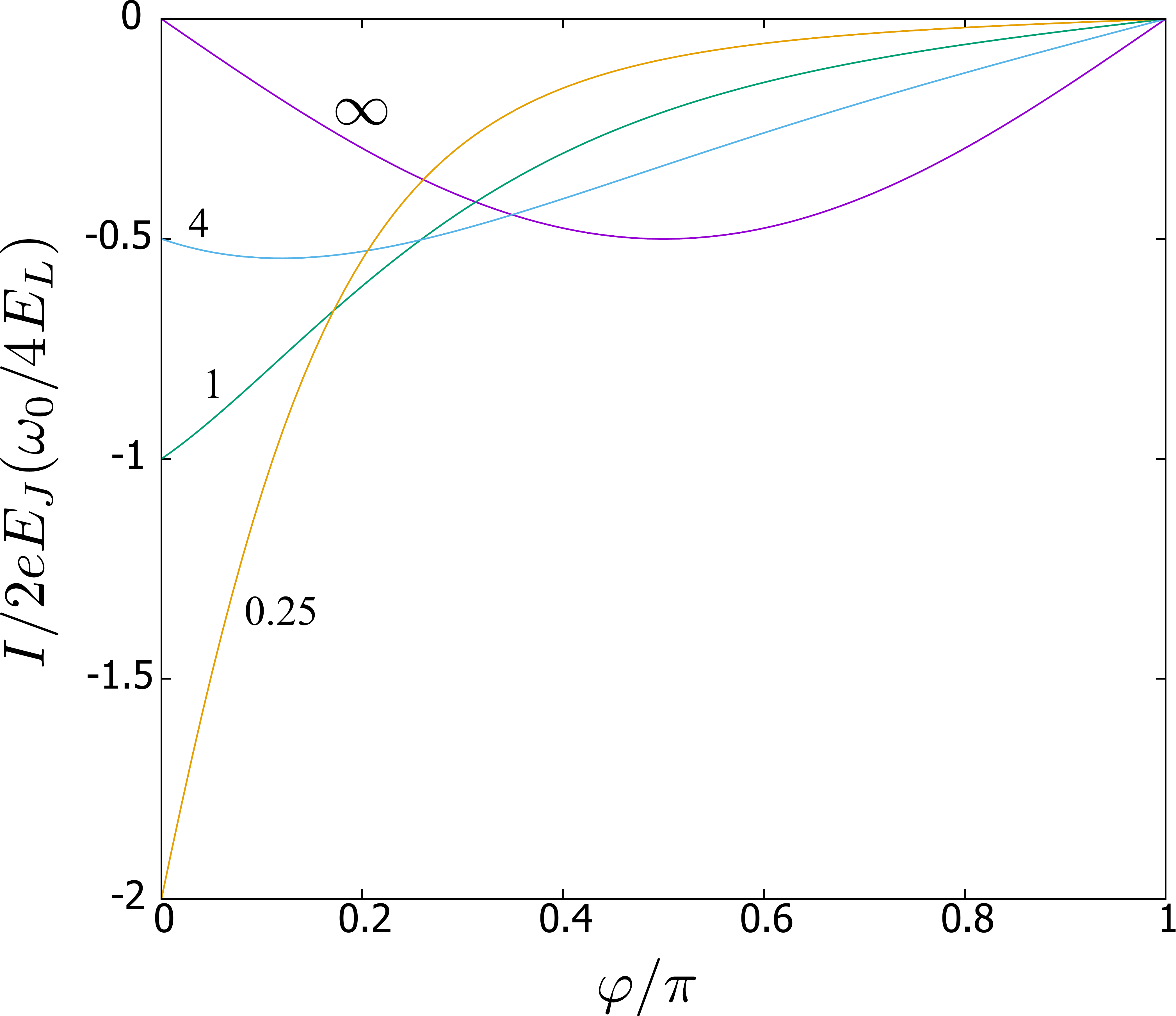}
\caption{\label{fig:currentosc} Small impedance, single oscillator case. The superconducting current in the odd-parity single-channel case versus phase for several values of $\zeta={\omega_0}/{(2 E_J)}$ corresponding to Eq.~\eqref{eq:single-osc-energy}.}
\end{figure}

We plot the phase dependence of the current for several values of $\zeta$ in Fig.~\ref{fig:currentosc}.

\subsection{Ohmic impedance} Let us now turn to the more complex case of an ohmic impedance. The impedance model in use comprises a capacitance, a resistor, and an inductance connected in parallel.
The resulting admittance at imaginary frequency  reads:
\begin{equation}
|\omega| {\cal Y}(\omega) = C \omega^2 + \frac{\omega}{R} + 1/L,
\end{equation} 
corresponding to the real-frequency admittance $Y(\omega) = - i\omega C + 1/R + 1/(-i\omega L)$.
The capacitance is here to cut the ohmic part at high frequencies $\omega_H = 1/RC$. It is important to understand that, in order to realize of a good phase bias at zero frequency, there must be a sufficiently small inductance here, $E_L \gg E^*_J$. This cuts the ohmic part at low frequency $\omega_L = R/L$. 

We concentrate on the case of small impedance $R e^2 \ll 1$. In this case, we can assume small fluctuations, $\varphi(\tau) = \varphi + f(\tau)$, $f(\tau) \ll \pi$, and expand similar to Eqs. \eqref{eq:expand1}, \eqref{eq:expand2} replacing $\hat{f} \rightarrow f(\tau)$. The spectrum of the fluctuations is given by
\begin{equation}
\langle |f_\omega|^2\rangle = 4 e^2 {\cal Z}(\omega)/|\omega|.
\end{equation}
Let us first compute the correction to the ground state energy of the even parity state. We expand $\cos(\varphi(\tau)) = \cos(\varphi) - \sin(\varphi) f(\tau) -\cos(\varphi) f^2(\tau)/2$. The correction consists of two terms: the first one is the first-order correction $\simeq f^2$, the second one is the second-order correction in $f$ at zero frequency -- it comes from the current induced in the inductance. It also causes a shift in $f$, $\delta f = - (E^*_J/ E_L) \sin \varphi $ . As a result, we find
\begin{eqnarray}
\label{eq:evenparitycorrection}
\delta E_g^{(e)} &=& E^*_J \cos(\varphi) \frac{\langle f^2\rangle}{2} - \frac{E_J^{*2}}{2 E_L} \sin^2\varphi \nonumber\\
&=& 
E^*_J  \cos(\varphi) \frac{2 e^2}{\pi} \int_0^\infty \frac{d\omega}{\omega} Z(\omega) - \frac{E_J^{*2}}{2 E_L} \sin^2\varphi \nonumber\\
&=& 
\alpha \ln \left[\frac{\omega_H}{\omega_L}\right] E^*_J \cos(\varphi) - \frac{E_J^{*2}}{2 E_L} \sin^2\varphi  
\nonumber\\
&=& 
- 2 \alpha \ln \left[\frac{\omega_H}{\omega_L}\right] E^*_J \sin^2\frac{\varphi}{2}  - \frac{E_J^{*2}}{2 E_L} \sin^2\varphi + {\rm const},
\end{eqnarray}
where we define the dimensionless impedance as $\alpha \equiv 2 R e^2/\pi$.
The resulting answer is somewhat similar to Eq.~\eqref{eq:sosceven}: the first term describes a small renormalization of $E^*_J$, while the second term gives the inductive energy.

Let us address the corrections to the ground state energy at odd parity.  For concreteness, we assume $s=1$, $\sin(\varphi/2)>0$. We expand $A$ in small deviations:
\begin{equation}
A = \sqrt{2 E_J} \left( \sin \frac{\varphi}{2} -\cos \frac{\varphi}{2} \frac{f(\tau)}{2} -\sin\frac{\varphi}{2} \frac{f^2(\tau)}{8}\right).
\end{equation}
Since we work in the lowest order in fluctuations, we need the self-energy part in the second order only, and the self-consistency equation reads
\begin{equation}
\label{eq:uptosecond}
\sqrt{\Omega} = \langle A \rangle + \Sigma^{(2)}(-i \Omega); \qquad \Sigma^{(2)}(-i \Omega) = \int \frac{d \omega}{2\pi} \frac{\langle\langle |A|^2_\omega\rangle \rangle}{\sqrt{\Omega + i \omega} - \langle A\rangle}.
\end{equation}
We expand it to the first order in the correction $\delta \Omega$, $\Omega_0 \equiv 2 E_J \sin^2(\varphi/2)$ being the binding energy in zeroth order.  This yields
\begin{equation}
\frac{\delta \Omega}{2\sqrt{\Omega}_0} = \delta \langle A \rangle + \frac{E_J}{2} \cos^2 \frac{\varphi}{2} \int \frac{d \omega}{2\pi} \frac{ 4 e^2 Z(\omega)}{ |\omega| \left( \sqrt{\Omega_0 + i \omega} - \sqrt{\Omega_0}\right)}; \qquad  \delta \langle A \rangle = -\frac{\sqrt{2 E_J}}{8} \sin \frac{\varphi}{2} \langle \langle f^2 \rangle \rangle 
+\frac{\sqrt{2 E_J}}{2} \cos \frac{\varphi}{2} \delta f .
\end{equation}
It is instructive to identify two parts in this correction. One is proportional to $\delta \langle A \rangle$, and we will call it coherent. The second part is proportional to  $\Sigma^{(2)}(-i \Omega)$ and is called incoherent. The coherent correction presents a renormalization of $E_J$ and a part of the inductive energy,
\begin{equation}
\label{eq:coherentcorrection}
\delta \Omega_{\rm{coh}} = -\alpha \ln \left[\frac{\omega_H}{\omega_L} \right] E_J \sin^2 \frac{\varphi}{2} - \frac{E_J E^*_J}{E_L} \sin^2 \varphi.
\end{equation}
If we compare this with the renormalization of $E^*_J$ in the even parity state, we note that the renormalization in the odd parity sector is two times smaller: a fact that we will recall when addressing the arbitrary impedance case. 

The analysis of the incoherent correction is complicated by the fact that it depends both on $\omega_L$ and $\Omega_0$. The upper cut-off is not relevant since the integral defining the correction converges at the maximum of these two frequencies.  If the impedance is not small, $\alpha \simeq 1$, $\omega_L$ surely exceeds $\Omega_0$ as  by definition of phase bias $E_L \gg E_J$. If the impedance is small, $\omega_L/\Omega_0 \simeq \alpha/(E_J/E_L)$ is the ratio of two small numbers and can be large or small. We give the general formulas below.

For our impedance model, 
\begin{equation}
\Sigma^{(2)}(- i\Omega) = \frac{E_J}{2} \cos^2 \frac{\varphi}{2} \int \frac{d \omega}{2\pi} \frac{ 4 e^2 R}{ \left(|\omega|+\omega_L\right) \left(\sqrt{\Omega + i \omega} - \sqrt{\Omega_0}\right) }.
\end{equation}
This is a tricky integral: the denominator contains a delta-functional contribution that is easy to miss at $\Omega \to \Omega_0$. To see this, let us transform the denominator assuming $\Omega \to \Omega_0$, $\omega \ll \Omega_0$ 
\begin{equation}
\frac{1}{\sqrt{\Omega + i \omega} - \sqrt{\Omega_0}} = \frac{\sqrt{\Omega + i \omega} + \sqrt{\Omega_0}}{\Omega-\Omega_0 + i \omega} \approx \frac{2\sqrt{\Omega_0}}{\Omega-\Omega_0 + i \omega} \to 2\pi\sqrt{\Omega}_0 {\rm sgn}(\delta \Omega) \delta(\omega).
\end{equation}
With this,
\begin{equation}
\Sigma^{(2)}(- i\Omega) = {E_J} \cos^2 \frac{\varphi}{2} \sqrt{\Omega_0}
2 e^2 L + 
  \frac{E_J}{2} \cos^2 \frac{\varphi}{2}\; {\rm p.v.}\int \frac{d \omega}{2\pi} \frac{ 4 e^2 R}{ \left(|\omega|+\omega_L\right) \left(\sqrt{\Omega_0 + i \omega} - \sqrt{\Omega_0}\right) },
\end{equation}
where p.v. indicates the principal-part integration. 
This gives the following correction:
\begin{equation}
\label{eq:incoherentcorrection}
\delta \Omega_{\rm{incoh}} =  \frac{E^2_J}{2 E_L} \sin^2\varphi + \alpha E_J \cos^2 \frac{\varphi}{2} X(x); \qquad x \equiv \frac{\omega_L}{2 E_J \sin^2 \frac{\varphi}{2}}.
\end{equation}
$X(x)$ is an elementary but rather complex function of $x$. Its asymptotes are
\begin{equation}
X(x) \approx \ln\left(\frac{4e}{x} \right) \quad{\rm at}\quad x \to 0; \qquad X(x) \approx \pi \sqrt{\frac{2}{x}} \quad{\rm at}\quad x \to \infty.
\end{equation} 
A compact expression is as follows $(z>1)$:
\begin{equation}
X((z^2-z^{-2})/2) = 4 z \left(\frac{{\rm arccoth}(z)}{z^2+1}+\frac{{\rm arccot}(1/z)}{z^2-1}-\frac{\pi z}{2(z^4-1)}\right).
\end{equation}

The correction to the ground state energy at odd parity defines the superconducting current that survives poisoning. We concentrate on the single-level case, $E_J = E^*_J$. In this case, the parts of theinductive energy in Eqs.~\eqref{eq:evenparitycorrection}, \eqref{eq:coherentcorrection}, \eqref{eq:incoherentcorrection} cancel each other as they did in the previous subsection, and the answer reads
\begin{equation}
\delta E_g^{(o)} = \delta E_g^{(e)} - \delta \Omega = -\alpha E_J\left( {\rm ln} \left[\frac{\omega_H}{\omega_L}\right] \sin^2 \frac{\varphi}{2} + \cos^2 \frac{\varphi}{2}  X\left(\frac{\omega_L}{2 E_J \sin^2 \frac{\varphi}{2}}\right)\right).
\end{equation}
Let us note that the first term gives a {\it negative} supercurrent. If it dominates (this is the case if the logarithmic factor is really bigger than 1), the Josephson junction becomes a $\pi$ junction. As for the second term, the energy correction is negative and reaches 0 both at $\varphi=0$ and $\varphi=\pi$. So if the second term is sufficiently big in comparison with the first one, there are two equivalent energy minima: one at $\varphi_0, 0<\varphi_0 <\pi$, and the symmetric one at $2\pi - \varphi_0$.

We observe the current jump at $\varphi \to 0$. In this case, $x \to \infty$ irrespective the ratio $\omega_L/E_J$  and the value of half-jump is given by
\begin{equation}
\label{eq:jumpOhm}
\frac{I_{{\rm hj}}}{2e} = - \pi \alpha E_J \sqrt{\frac{E_J}{\omega_L}} = - \frac{ E_J}{2} \sqrt{\frac{E_J}{E_L}} \sqrt{\frac{\omega_L}{E_L}}.  
\end{equation}
Note the similarity with Eq.~\eqref{eq:currentjump} if associating $\omega_L$ and $\omega_0/\sqrt{2}$. 

Let us look at the limiting cases. If $\omega_L \gg E_J$, the argument of $X$ is always bigger than 1 and the limiting form of the correction reads
\begin{equation}
\delta E_g^{(o)} = \alpha E_J\left( - {\rm ln} \left[\frac{\omega_H}{\omega_L}\right] \sin^2 \frac{\varphi}{2} -  \pi \sqrt{\frac{E_J}{\omega_L}} \cos \frac{\varphi}{2} \sin \varphi \right).
\end{equation}
The first term dominates over the whole range of $\varphi$ except $\varphi \to 0$, the second one gives rise to a jump at $\varphi \to 0$.

In the opposite case, $\omega_L \ll E_J$, the argument of $X$ is swept from large to small values within a narrow interval of the phase, $\varphi \simeq \sqrt{\omega_L/E_J}$. The current is concentrated in this interval,

\begin{equation}
I(\varphi) = - |I_{\rm hj}| f(\varphi/\sqrt{2\omega_L/E_J}); \qquad f(x) \equiv - \frac{\sqrt{2}}{\pi} X'(x^{-2})/x^3\enspace{\rm with}\enspace f(0)=1, \enspace f(x) \to \sqrt{2}/(\pi x) \enspace {\rm at} \enspace x \to \infty.   
\end{equation}
Note again some similarities with Eq.~\eqref{eq:narrowcurrent}: The current scales with $I_{\rm hj}$ and is concentrated at small $\varphi$, though the scaling functions are different. 

\begin{figure}
\includegraphics[width=0.5\columnwidth]{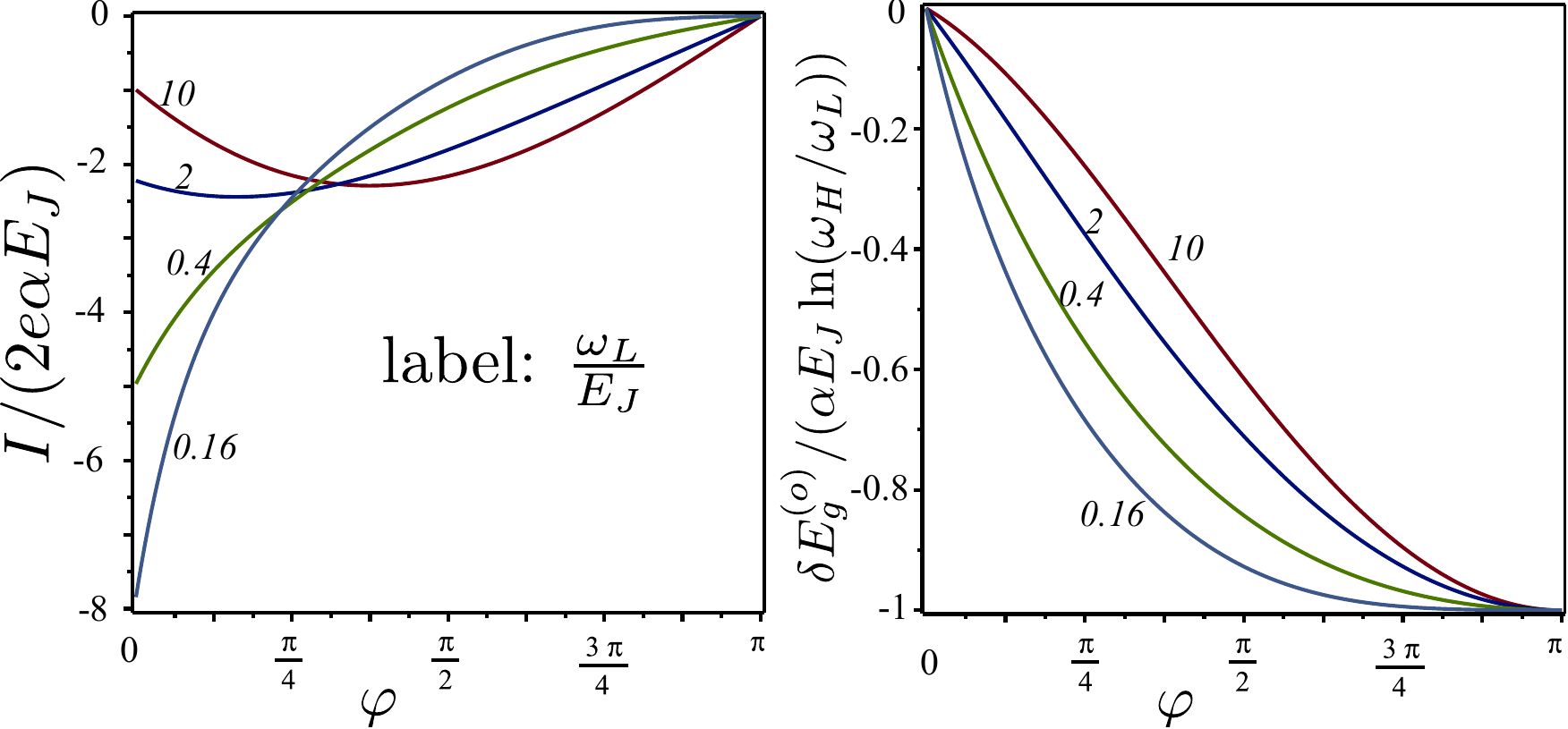}
\caption{\label{fig:current} Small ohmic impedance. Left: The superconducting current in the odd-parity single-channel case versus phase. Right: corresponding energies. We set $\ln(\omega_H/\omega_L) =5$, the curve labels correspond to various values of $\omega_L/E_J$. } 
\end{figure}

We plot the resulting current in Fig.~\ref{fig:current} for $\omega_L/E_J = 10,2,0.4,0.16$. For this example, we set $\ln(\omega_H/\omega_L) =5$. In this case, the energy minimum is at $\varphi=\pi$ for all $\omega_L/E_J$ in use.

\subsection{Beyond perturbation theory}

It follows from the previous analysis that $\delta \Omega \propto \varphi$ at $\varphi \to 0$ while $\Omega_0 \propto \varphi^2$. Therefore, $\delta \Omega$ becomes comparable with $\Omega_0$ at sufficiently small $\varphi$ and the perturbation theory should not work. However, there is a simple workaround common for both small-impedance models we consider.

To start with, let us understand when it is plausible to disregard the higher-order diagrams in the self-energy.
The second-order $\Sigma^{(2)}$ can be estimated as $E_J \langle f^2\rangle/\sqrt{{\rm max} (\Omega, \omega)}$, where $\omega$ here is either $\omega_0$ or $\omega_L$ from previous subsections. The fourth-order $\Sigma^{(4)}$ contains 3 $G_0$ and two A-correlators, therefore it can be estimated as $E_J^2 (\langle f^2\rangle)^2/ (\sqrt{{\rm max} (\Omega, \omega)})^3$. They differ by a factor $E_J \langle f^2\rangle/{\rm max} (\Omega, \omega)$ that has to be small to neglect higher orders. (The estimation of the relative value of $\Sigma ^{(3)}$ and all odd order diagramms is even smaller since we consider $\varphi \ll \pi$.)

Let us now assume $\Omega \ll \omega$ at $\varphi=0$. Then $\Sigma^{(2)}$ does not depend on $\varphi$, and $\Omega$ can be estimated as $\Omega \simeq (\Sigma^{(2)})^2 \simeq E_J^2 (\langle f^2\rangle)^2 /\omega$. 
Since $\langle f^2\rangle \simeq \omega/E_L$, one obtains $\Omega/\omega \simeq (E_J/E_L)^2 \ll 1$ under the assumption of phase bias. 
The higher orders can be neglected if $E_J \langle f^2\rangle/\omega \simeq E_J/E_L \ll 1$, that is, under the same assumption.

The phase dependence is incorporated through $\langle A \rangle = \sqrt{E_J/2}\, s \varphi$. The resulting expression for $\Omega$ can be presented in the following form:
\begin{equation}
\label{eq:smallphistructure}
\sqrt{\Omega} = \sqrt{E_J/2} \left( s \varphi + \varphi_c\right)
\end{equation} 
with $\varphi_c \simeq \sqrt{E_J/E_L} \sqrt{\omega/E_L}$. The value of $\varphi_c$ can be also obtained from Eqs. \eqref{eq:currentjump}, \eqref{eq:jumpOhm}, 
$\varphi_c = (|I_{\rm hj}|/2e)/E_J$.

\begin{figure}
\includegraphics[width=0.4\columnwidth]{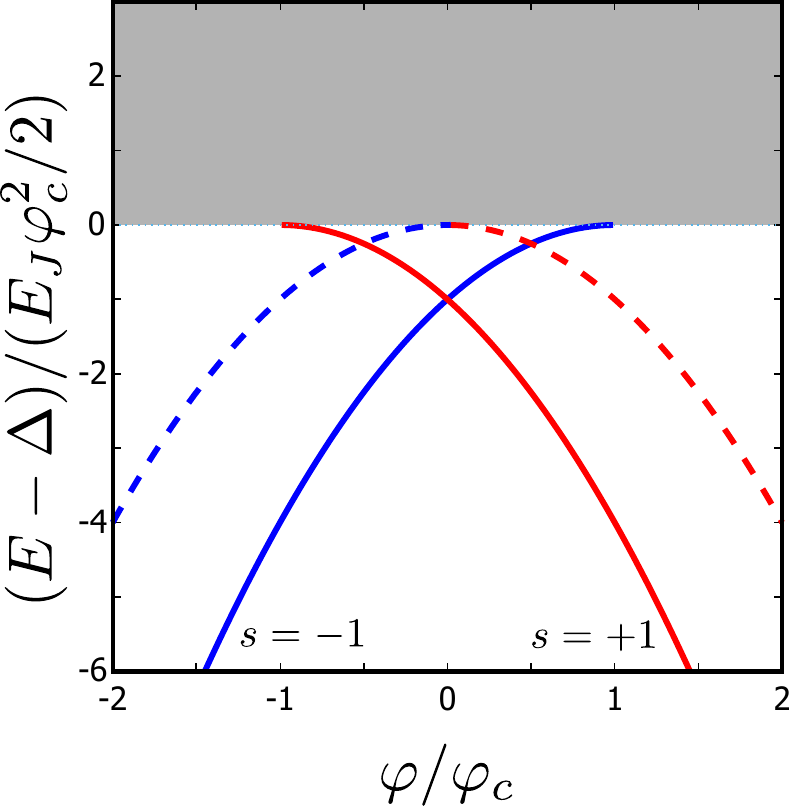}
\caption{\label{fig:smallphi} Bound states in the narrow interval of the phases for $s=\pm 1$ (blue and red). Thin curves: no interaction.}
\end{figure}

We note that in the narrow interval of the phase $-\varphi_c < \varphi < \varphi_c$ the bound state is present for both values of $s$ (see Fig.~\ref{fig:smallphi}): a qualitative change as compared to the situation without interaction. 

\section{Arbitrary impedance: phase bias}
In this Section, we address larger impedances $\alpha \simeq 1$. It is essential that we keep the conditions of phase bias, $E_L \gg E^*_J, E_J$. In this case, the low cut-off frequency $\omega_L$ is much bigger than $E^*_J, E_J$ and does not change upon renormalization of $E_J$. This renormalization is thus always finite. This  implies that there is no Schmid transition  at phase bias, contrary to what happens at  {\em current} bias. This situation will be addressed in the next Section.

For the consideration, we will need the averages of the phase exponents over the fluctuations induced by the environment: let us cite these results here ($\beta$ is arbitrary at the moment, we will need $\beta =1, 1/2$):
\begin{eqnarray}
\langle e^{- i \beta \varphi(\tau)} \rangle &=& e^{-i \beta (\varphi + \langle f \rangle)} e^{-\frac{\beta^2}{2} \langle\hspace{-2pt}\langle f^2\rangle\hspace{-2pt}\rangle} ;\qquad \langle\hspace{-2pt}\langle f^2\rangle\hspace{-2pt}\rangle = \int \frac{d\omega}{2\pi} \frac{4e^2{\cal Z}(\omega)}{|\omega|};\qquad 
\langle e^{- i \beta (\varphi(0)-\varphi(\tau))}\rangle =
e^{-\frac{\beta^2}{2} U(\tau)};\\
U(\tau) &\equiv& \langle (f(0)-f(\tau))^2\rangle = 4 \int \frac{d\omega}{2\pi} \sin^2(\omega\tau/2) \frac{4e^2{\cal Z}(\omega)}{|\omega|} \quad {\rm with} \quad
U(0) = 0, \enspace U(\tau \to \infty) = 2 \langle f^2\rangle ; \\
\langle e^{- i \beta (\varphi(0)+\varphi(\tau))}\rangle &=&
e^{-i 2 \beta (\varphi+ \langle f \rangle)} e^{\frac{\beta^2}{2} U(\tau) - 2 \beta^2 \langle\hspace{-2pt}\langle f^2\rangle\hspace{-2pt}\rangle}.
\end{eqnarray}

In the leading approximation, $\langle\hspace{-2pt}\langle f^2\rangle\hspace{-2pt}\rangle = 2 \alpha \ln(\omega_H/\omega_L)$ and $U(\tau) = 4 \alpha \ln(\omega_H \tau)$ at $\omega_H^{-1} \ll \tau \ll \omega_L^{-1}$. One needs to use these formulas with care since the ratio $\omega_H/ \omega_L \sim (E_C/E_L) \alpha^{-2}$ depends on $\alpha$ as well.

Let us address the even parity sector first. The phase-dependent part of the energy is given by the renormalized (and strongly suppressed) Josephson energy,
\begin{equation}
E_g^{(e),1} = - \tilde{E}^*_J \cos \varphi \enspace{\rm with}\enspace \frac{\tilde{E}^*_J}{E^*_J}= e^{-\langle\hspace{-2pt}\langle f^2\rangle\hspace{-2pt}\rangle /2} \simeq \left[ \frac{\omega_L}{ \omega_H}\right]^\alpha.
\label{eq-sup-even}
\end{equation}
There is also a second-order contribution that is not exponentially suppressed. 
\begin{equation}
E_g^{(e),2} = - E^*_J \int d\tau \left[e^{-U(\tau)/2}- e^{-\langle\hspace{-2pt}\langle f^2\rangle\hspace{-2pt}\rangle }\right].
\end{equation}
For strong suppression ($\tilde E_J^*/E_J^*\ll1$), the integrand at $\omega_H^{-1} \ll \tau \ll \omega_L^{-1}$ is $(\omega_H \tau)^{-2\alpha}$.
Thus, if $\alpha > 1/2$, the integral converges at the upper cut-off $\omega_H \tau \simeq 1$ whereas it converges at the lower cut-off otherwise. So we estimate (skipping the prefactors):
\begin{align}
E_g^{(e),2} \simeq \frac{E^{*2}_J}{\omega_H} \ \ \ {\rm if} \ \alpha>1/2 ; \qquad
E_g^{(e),2} \simeq \frac{E^{*2}_J}{\omega_L} \left[ \frac{\omega_L}{ \omega_H}\right]^{2\alpha} \simeq \frac{\tilde{E}^{*2}_J}{\omega_L}\ \ \ {\rm if} \ \alpha<1/2.
\end{align}
The correction at $\alpha<1/2$ is related to the one discussed in Ref.~\cite{Hekking1997}.  We conclude that the second-order contribution dominates at $\alpha>1/2$.

We turn to odd parity and keep  terms up to the second order  in the self-consistency  equation (cf. Eq.~\eqref{eq:uptosecond}),
\begin{equation}
\sqrt{\Omega} = \langle A \rangle + \Sigma^{(2)}(-i \Omega).
\end{equation}
The average $A$ is strongly suppressed,
\begin{equation}
\langle A \rangle = \sqrt{2 \tilde{E}_J} s \sin \frac\varphi2 \enspace {\rm with} \enspace \frac{\tilde{E}_J}{E_J} = e^{- {\langle f^2 \rangle}/{4}} \simeq \left[ \frac{\omega_L}{ \omega_H}\right]^{\alpha/2}.
\label{eq-sup-odd}
\end{equation}
As was noted when considering the small impedance limit, the suppression of the odd-parity Josephson coupling is two times weaker than the one of the even-parity Josephson coupling (see Eq.~\eqref{eq-sup-even}).

At $\alpha <1$, $\Sigma^{(2)}$ can be neglected in zeroth approximation. The superconducting current is given by
the renormalized expression
\begin{equation}
\frac{I(\varphi)}{2e} = (\tilde{E}^*_J - \tilde{E}_J) \sin\varphi = \left( E^*_J e^{-\frac{\langle f^2\rangle}{2}} - E_J e^{-\frac{\langle f^2\rangle}{4}}\right) \sin\varphi.
\end{equation} 
Since $E_J$ is less suppressed than $E_J^*$, there is a chance that the supercurrent in the odd state is bigger in magnitude than that in the even state. The current at phases $\varphi\in[0,\pi]$ is negative in this case. For the single-channel case $E^*_J=E_J$ the current is always negative in this phase interval.

\begin{figure}
\includegraphics[width=0.8\columnwidth]{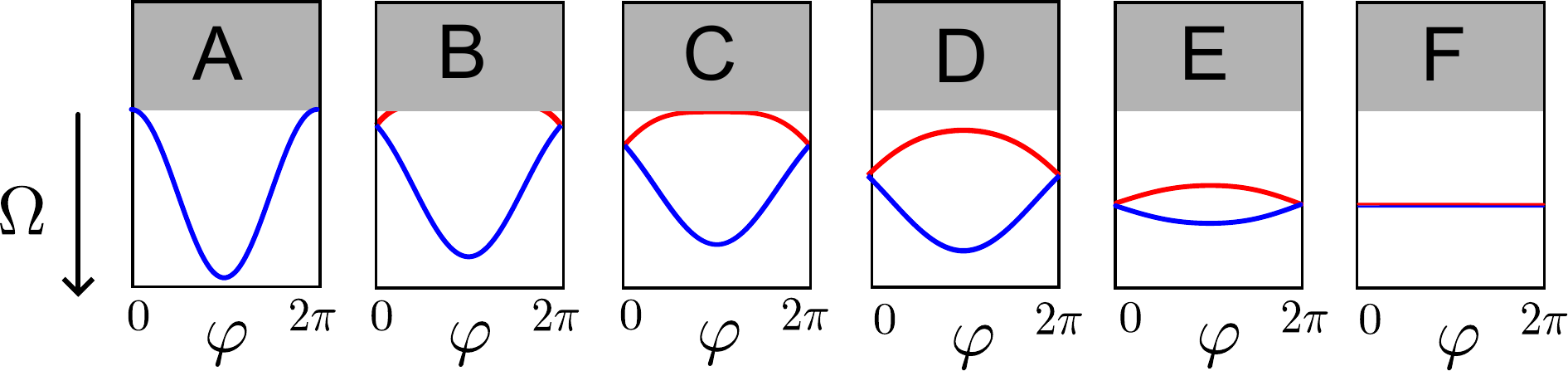}
\caption{\label{fig:levels} The various bound regimes that can occur in the odd-parity Josephson junction. Qualitatively, they can be described with Eq.~\eqref{eq:boundregimes}.  A: only one superposition gives rise to a bound states (realized at no interaction, $\alpha=0$). B: two bound states in a finite phase interval near $\varphi=0,2\pi$. (cf.~Fig \ref{fig:smallphi}) C: transition between B and D. D: both bound states are present at all phases, the resulting superconducting current is $4 \pi$ periodic. E: the splitting of two bound states is much smaller than their average phase-independent energy. F: Two bound states are degenerate, and bear no phase-dependence. }
\end{figure}

However, the second order term $\Sigma^{(2)}(-i \Omega)$ can become important since it has a phase-independent part that controls the position of the bound state with respect to the continuum edge.  This leads to a variety of {\it bound regimes} A-F listed in Fig.~\ref{fig:levels}. 
To estimate $\Sigma^{(2)}$, we can concentrate on the phase-independent terms  that are not  exponentially suppressed at large $\alpha$ and, since $\Omega \ll \omega_L$, disregard their $\Omega$ dependence. This yields
\begin{equation}
\Sigma^{(2)} = {E_J} \int_0^{\infty} \frac{d \tau}{\sqrt{\pi \tau}} \langle\hspace{-2pt}\langle e^{i \varphi(0)/2} e^{-i \varphi(\tau)/2}\rangle\hspace{-2pt}\rangle.
\end{equation}
The integrand at $\omega_H^{-1} \ll \tau \ll \omega_L^{-1}$ is $(\omega_H \tau)^{-\alpha/2} \tau^{-1/2}$. The integral thus converges at the upper cut-off if $\alpha >1$ and at the lower cut-off if $\alpha <1$. The estimations for $\Sigma^{(2)}$ then read
\begin{align}
\label{eq:sigmaest1}
\Sigma^{(2)} \simeq \frac{E_J}{\sqrt{\omega_H}} \ \ \ {\rm if} \ \alpha>1; \;\\
\label{eq:sigmaest}
\Sigma^{(2)} \simeq \frac{E_J}{\sqrt{\omega_L}} 
\left[ \frac{\omega_L}{ \omega_H}\right]^{\alpha/2} \simeq \frac{\tilde{E}_J}{\sqrt{\omega_L}}\ \ \ {\rm if} \ \alpha<1.
\end{align}

If $\langle A \rangle$ dominates but the finite $\Sigma^{(2)}$ is taken into account, there is a small current jump at $\varphi \to 0$, namely $|I_{\rm hj}|/(2e) \simeq \tilde{E_J} \sqrt{\tilde E_J/\omega_L}$ corresponding to the level structure described by Eq.~\eqref{eq:smallphistructure} with $\varphi_c = \Sigma^{(2)}/\sqrt{\tilde{E}_J/2} \simeq \sqrt{\tilde E_J/\omega_L}$ (regime B). Eventually, $\varphi_c$ and the relative value of the jump $|I_{\rm hj}|/(2e\tilde E_J)$ decrease with increasing $\alpha$ at $1/\ln(\omega_H/\omega_L)<\alpha<1$.

We compare $\langle A \rangle$ and $\Sigma^{(2)}$ to establish that the latter dominates at $\alpha > 2(1+ \ln(\omega_L/E_J)/\ln(\omega_H/\omega_L))\equiv \alpha_c >2$.  At $\alpha_c > \alpha >1$ we are still in the regime B, with the only difference that $ \Sigma^{(2)}$ saturates at the value $\simeq E_J/\sqrt{\omega_H}$ and therefore $\varphi_c$ and  the relative current jump {\it increase} with increasing $\alpha$, namely $\varphi_c \simeq E_J/\sqrt{\omega_H \tilde{E_J}}$.

At $\alpha \approx \alpha_c$, where $\sqrt{2 \tilde{E}_J} = \Sigma^{(2)}$, an important transition (regime C) takes place: the bound state is present  at any phase for both $s=\pm 1$ (regime D). As a consequence, the odd parity state becomes stable upon an adiabatic sweep of the phase. The bound  state energies are given by 
\begin{equation}
\label{eq:boundregimes}
\Omega = \left(s \sqrt{2 \tilde{E}_J} \sin \frac{\varphi}{2} +  \Sigma^{(2)}\right)^2.
\end{equation}
The resulting superconducting current at a given $s$ becomes $4\pi$ periodic, a phenomenon similar to that signifying the presence of Majorana modes \cite{Mayorana4pi}. The $2\pi$ periodicity is restored upon relaxation to the lowest energy state  within the odd sector.

At $\alpha > \alpha_c$, the bound state energy $\Omega \simeq E_J^2/\omega_H$ hardly depends on the phase and $\alpha$. The remaining phase dependence results in a strongly suppressed supercurrent (regime E). 
\begin{equation}
\label{eq:almostdegenerate}
I(\varphi)/e \sim E_J\sqrt{\frac{\tilde{E}_J}{\omega_H}}
s \cos \frac{\varphi}{2}.
\end{equation}
Despite being suppressed, this supercurrent is parametrically bigger than the one $\propto \tilde E_J^*$ in the even parity state.

The resulting supercurrent in this regime is $4\pi$ periodic provided $s$ is conserved during the measurement time. The relaxation to the energetically favourable value of $s$ would restore $2\pi$ periodicity. More involved research is required to estimate a typical relaxation time in this situation: let us explain why.  The terms in the Hamiltonian that break the conservation of $s$ requires an asymmetry of the quasiparticle wave function in the right/left lead. Such asymmetry is not manifested under standard assumptions of semiclassical theory of superconductivity, where the difference between the leads effectively averages out. The asymmetry may arise owing to disorder-induced fluctuations of the superconducting order parameter \cite{LarkinDisorder} or mesoscopic fluctuations of Andreev scattering \cite{Feigelman}. This provides a small factor associated with the ratio of the electronic wavelength and the spread of an Andreev state over the leads. In addition to this symmetry violation, the relaxation requires an inelastic processes to bridge the energy difference between the split states with different $s$. This inelastic process may arise from the electromagnetic environment. With this, a typical relaxation time may be even longer than the life time of the odd parity state under consideration.

\section{Arbitrary impedance: current bias}

In contrast with the phase bias situation, there is no built-in low-energy cut-off at current bias: the lower cut-off is determined by the renormalized Josephson energy.

Let us remind how this works by addressing the even parity sector. The renormalized $\tilde{E}^*_J$ is given by the same formula as in the previous section,  see Eq.~\eqref{eq-sup-even},
where $\omega_L$ is estimated as $\tilde{E}^*_J$. (This estimation is fine in the region of interest $\alpha \simeq 1$. At $\alpha \ll 1$, $\tilde{E}^*_J \alpha$ would be a more accurate approximation. Yet taking into account the difference between these two estimations exceeds the accuracy of the approach.) With this,
\begin{equation}
\frac{\tilde{E}^*_J}{E^*_J} = \left[ \frac{E^*_J}{ \omega_H}\right]^\frac{\alpha}{1-\alpha},
\end{equation}
that is, $\tilde E_J^*$ vanishes at the Schmid transition, $\alpha=1$.  

Let us now turn to the odd parity sector and assume $E^*_J = E_J$. 
(The analysis of a many-channel situation $E^{*}_J \gg E_J$ leads to a more involved situation where both $\tilde{E}^*_J$ and $\tilde{E}_J$ may play a role of the low-energy cut-off. This analysis is beyond the scope of the present paper.)

To start with, let us concentrate on the interval $\alpha<1$. In this case, the lower cut-off can be unambiguously identified as $\tilde{E}_J$. Applying the results of the previous section (see Eq.~\eqref{eq-sup-odd}), we obtain
\begin{equation}
\frac{\tilde{E}_J}{E_J} \simeq \left[ \frac{E_J}{ \omega_H}\right]^{\alpha/2} \quad\to\quad \frac{\tilde{E}_J}{E_J} = \left[ \frac{E_J}{ \omega_H}\right]^\frac{\alpha}{2-\alpha}.
\end{equation}
The estimation of $\Sigma^{(2)}$ with the help of Eq.~\eqref{eq:sigmaest} gives $\Sigma^{(2)} \simeq \sqrt{\tilde{E}_J}$. We see that, by contrast to the situation of phase bias, the first- and second-order contributions are of the same order of magnitude, as well as all higher orders. Therefore the accuracy of the method does not allow to predict the phase dependence of the energy, nor if bound states persist for both values of $s$, as it was the case under phase bias (regimes B-C-D). However, we still note and use the difference in  the renormalizations for phase-dependent ($\tilde{E}_J$) and phase-independent parts of $\sqrt{\Omega}$.

This becomes important at $\alpha >1$, where in accordance with Eq.~\eqref{eq:sigmaest1} $\Sigma^{(2)}$ does not depend on the low cut-off anymore and saturates at the value $\simeq E_J/\sqrt{\omega_H}$. 
As to the phase-dependent part, it  further decreases with increasing $\alpha$. This brings us to regime E: almost degenerate bound states described by Eq.~\eqref{eq:almostdegenerate}. In this case, the lower cut-off is $\simeq E_J\sqrt{\frac{\tilde{E}_J}{\omega_H}}$ rather than $\simeq \tilde{E}_J$, which yields
\begin{equation}
 \frac{\tilde{E}_J}{E_J} = \left[ \frac{E_J}{ \omega_H}\right]^\frac{3\alpha}{4-\alpha}.
\end{equation} 
We see that $\tilde{E}_J$ vanishes at $\alpha=4$. This is the Schmid transition point for a half of the Cooper pair charge corresponding to a $4\pi$ periodicity in phase.

The bound state is completely degenerate with respect to $s$ at $\alpha >4$ (regime F). If we recall the spin, we observe the realization of a 4-fold degeneracy. 

\begin{figure}
\includegraphics[width=\columnwidth]{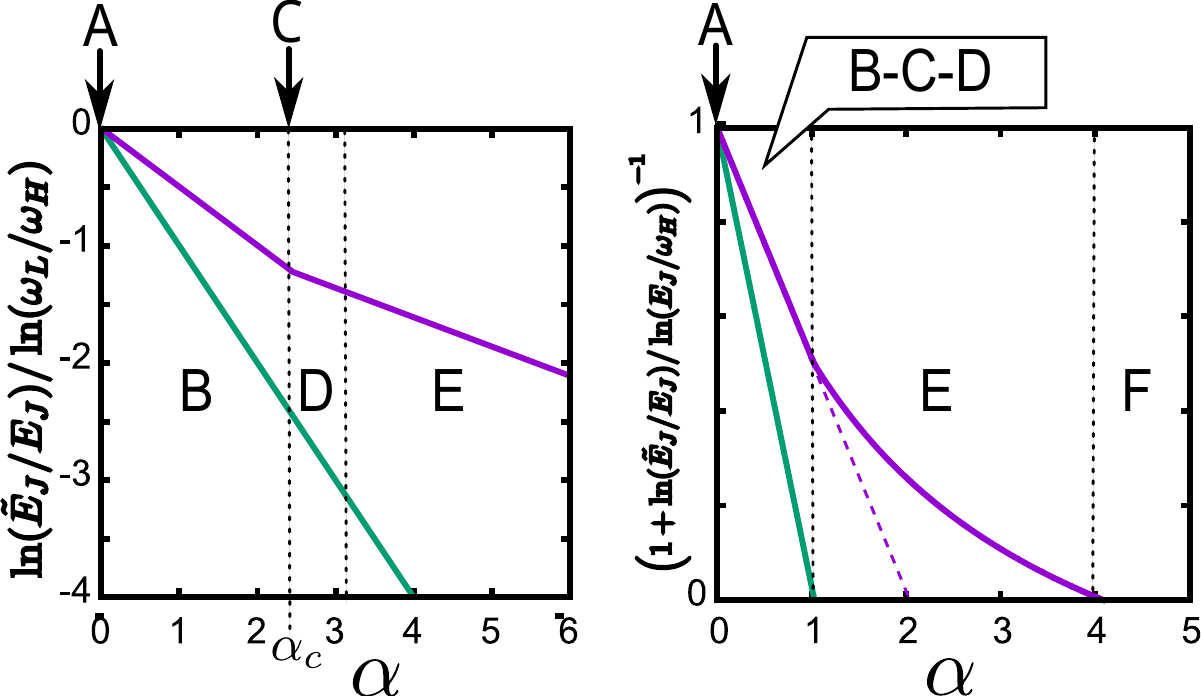}
\caption{\label{fig:arbitrary} The summary of the results at arbitrary impedance. Renormalized Josephson energies $\tilde{E}_J$ at both parities.  Capital letters indicate the bound regimes at odd parity, dotted vertical lines separate the regimes. Left: phase bias, $\tilde{E}_J$ never vanishes. The separating regime $C$ occurs at $\alpha=\alpha_c$. Right: current bias. Schmid transition is at $\alpha=1$ for even parity and at $\alpha=4$ for odd one. The renormalization law at odd parity changes at $\alpha=1$. }
\end{figure}

The results of the two last Sections are summarized in Fig.~\ref{fig:arbitrary}.

\bibliography{JosOdd-bib}
